\documentclass[12pt,twoside]{article}

\def\TheAuthor{J\"urgen Schnack}                                   
\def\TheTitle{Quantum Theory of Molecular Magnetism}                        
\def\TheHeading{Quantum Theory of Molecular Magnetism}                                      
\def\TheInstitute{Fachbereich Physik}       
\def\TheAddress{Universit\"at Osnabr\"uck}              
\def\ThePart{X}                                               
\def\TheChapter{5}                                            

\usepackage{ferienschule_04}                                  

\usepackage{multirow}
\usepackage{amssymb,amsbsy,amsmath,amscd,amsfonts}

\newcommand {\mofe} {\{$\textrm{Mo}_{72}\textrm{Fe}_{30}$\}}

\newcommand{\Lanczos}{Lanczos}
\newcommand{\Neel}{N\'{e}el}
\newcommand{\op}[1]{%
    \fontdimen12\textfont3=2pt\fontdimen12\scriptfont3=1.4pt%
    \!\null\mathop{\vphantom{#1}\smash{#1}}\limits_{\sim}\null\!}
\newcommand{\EinsOp}
           {\;\smash{\raisebox{-1.1ex}{$\!\!\stackrel{\!\mbox{1}
            \hspace{-0.4ex}\rule[0.0ex]{0.06ex}{1.60ex}}{\sim}$}}}
\newcommand{\xref}[1]{\protect\ref{#1}}
\newcommand{\fmref}[1]{(\protect\ref{#1})}
\def\bra#1{\langle \, {#1} \, | \,}
\def\ket#1{\, | \, {#1} \, \rangle}
\newcommand{\braket}[2]{\langle \, {#1} \, | \, {#2} \, \rangle}
\def\half{{\frac{1}{2}}}

\newcommand{\SMax}{S_{\mbox{\scriptsize max}}}

\newcommand{\pp}[2]{\frac{\partial \, {#1}}{\partial \, {#2}}}
\newcommand{\dd}[2]{\frac{{d}\, {#1}}{{d} {#2}}\;}
\newcommand{\ddt}{\frac{d}{dt}}
\newcommand{\dint}{\text{d}}

\newcommand{\tr}{\mbox{tr}}

\begin{document}

\MakeTitel           

\tableofcontents     

\newpage


\section{Introduction}
\label{magmol-sec:1}

The synthesis of molecular magnets has undergone rapid progress
in recent years
\cite{SGC:Nat93,GCR:S94,Gat:AM94,Cor:NATO96,MPP:CR98,CGS:JMMM99}.
Each of the identical molecular units can contain as few as two
and up to several dozens of paramagnetic ions (``spins"). One of
the largest paramagnetic molecules synthesized to date, the
polyoxometalate \{Mo$_{72}$Fe$_{30}$\} \cite{MSS:ACIE99}
contains 30 iron ions of spin $s=5/2$.  Although these materials
appear as macroscopic samples, i.~e. crystals or powders, the
intermolecular magnetic interactions are utterly negligible as
compared to the intramolecular interactions.  Therefore,
measurements of their magnetic properties reflect mainly
ensemble properties of single molecules.

Their magnetic features promise a variety of applications in
physics, magneto-chemistry, biology, biomedicine and material
sciences \cite{SGC:Nat93,Gat:AM94,Cor:NATO96} as well as in
quantum computing \cite{LeL:Nature01,WaZ:PLA02,Wang:PRA02A}. The
most promising progress so far is being made in the field of
spin crossover substances using effects like ``Light Induced
Excited Spin State Trapping (LIESST)" \cite{GHS:AC94}.

It appears that in the majority of these molecules the localized
single-particle magnetic moments couple antiferromagnetically
and the spectrum is rather well described by the Heisenberg
model with isotropic nearest neighbor interaction sometimes
augmented by anisotropy terms
\cite{BeG90,DGP:IC93,CCF:IC95,PDK:JMMM97,WSK:IO99}.  Thus, the
interest in the Heisenberg model, which is known already for a
long time \cite{Heisenberg:ZP28}, but used mostly for infinite
one-, two-, and three-dimensional systems, was renewed by the
successful synthesis of magnetic molecules. Studying such spin
arrays focuses on qualitatively new physics caused by the finite
size of the system.

Several problems can be solved with classical spin dynamics,
which turns out to provide accurate quantitative results for
static properties, such as magnetic susceptibility, down to
thermal energies of the order of the exchange coupling. However,
classical spin dynamics will not be the subject of this chapter,
it is covered in many publications on Monte-Carlo and
thermostated spin dynamics. One overview article which discusses
classical spin models in the context of spin glasses is given by
Ref.~\cite{BiY:RMP86}.

Theoretical inorganic chemistry itself provides several methods
to understand and describe molecular magnetism, see for instance
Ref.~\cite{Boca99}. In this chapter we would like to focus on
those subjects which are of general interest in the context of
this book.

\section{Substances}
\label{magmol-sec:2}

From the viewpoint of theoretical magnetism it is not so
important which chemical structures magnetic molecules actually
have. Nevertheless, it is very interesting to note that they
appear in almost all branches of chemistry. There are inorganic
magnetic molecules like polyoxometalates, metal-organic
molecules, and purely organic magnetic molecules in which
radicals carry the magnetic moments. It is also fascinating that
such molecules can be synthesized in a huge variety of
structures extending from rather unsymmetric structures to
highly symmetric rings.

\begin{figure}[ht!]
\centering
\includegraphics[height=45mm]{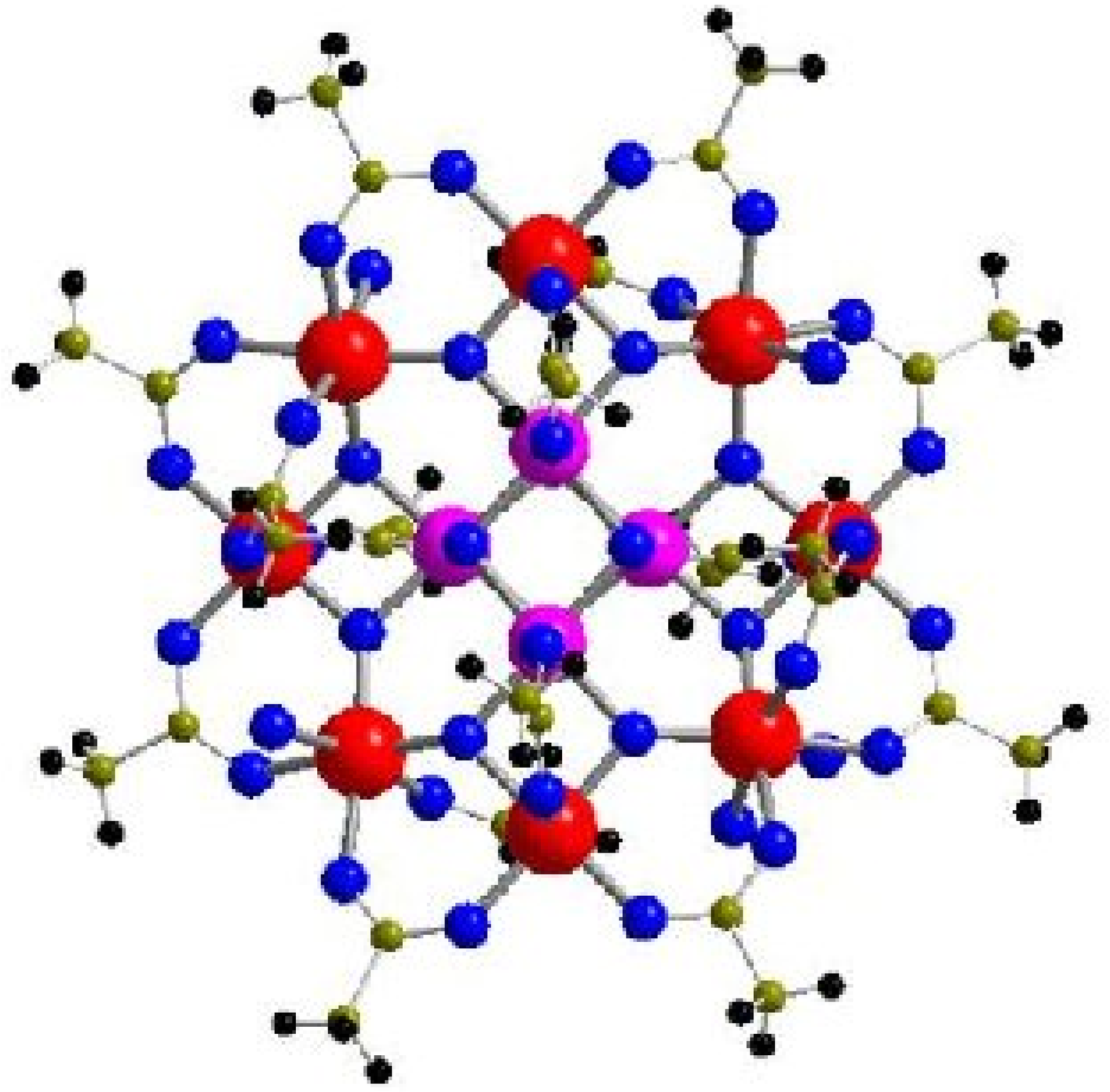}
$\qquad$
\includegraphics[height=45mm]{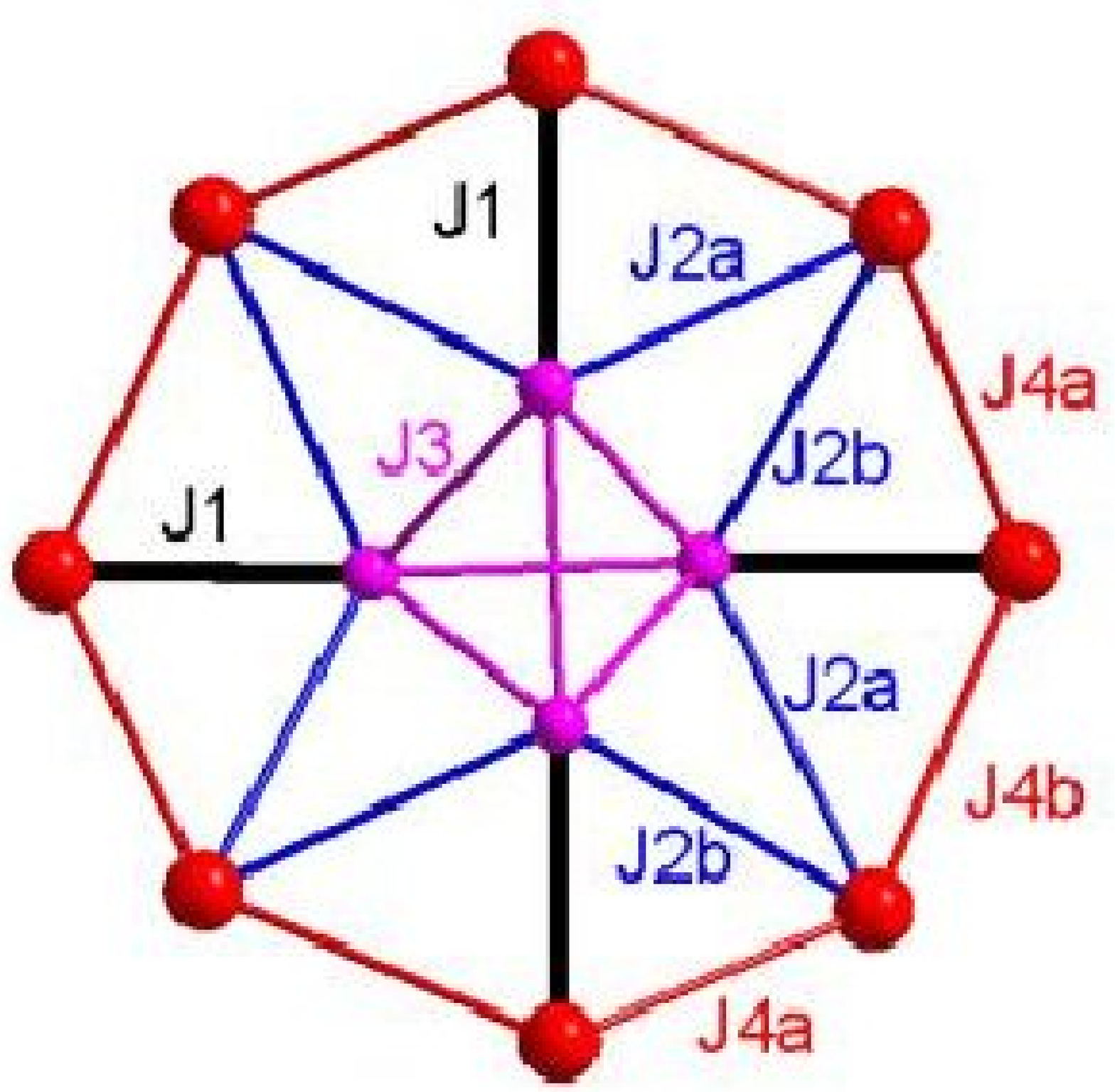}
\caption{Structure of Mn-12-acetate: On the l.h.s. the Mn ions
  are depicted by large spheres, on the r.h.s. the dominant
  couplings are given. With friendly permission by G.~Chaboussant.}
\label{magmol-fig:Mn12}
\end{figure}
One of the first magnetic molecules to be synthesized was
Mn-12-acetate \cite{Lis:ACB80} (Mn$_{12}$) --
\linebreak[4]
[Mn$_{12}$O$_{12}$(CH$_{3}$COO)$_{16}$(H$_{2}$O)$_{4}$] -- which
by now serves as the ``drosophila" of molecular magnetism, see
e.~g.
\cite{SGC:Nat93,TLB:Nature96,Cor:NATO96,JLB:PRL00,FWK:PRB01}.
As shown in Fig.~\ref{magmol-fig:Mn12} the molecules contains
four Mn(IV) ions ($s=3/2$) and eight Mn(III) ions ($s=2$) which
are magnetically coupled to give an $S=10$ ground state. The
molecules possesses a magnetic anisotropy, which determines the
observed relaxation of the magnetization and quantum
tunneling\index{magnetization tunneling}\index{tunneling} at low
temperatures \cite{TLB:Nature96,ThB:JLTP98}.

Although the investigation of magnetic molecules in general --
and of Mn-12-acetate in particular -- has made great advances
over the past two decades, it is still a challenge to deduce the
underlying microscopic Hamiltonian, even if the Hamiltonian is
of Heisenberg type. Mn-12-acetate is known for about 20 years
now and investigated like no other magnetic molecule, but only
recently its model parameters could be estimated with satisfying
accuracy \cite{RJS:PRB02,Honecker03}.

\begin{figure}[ht!]
\centering
\includegraphics[height=45mm]{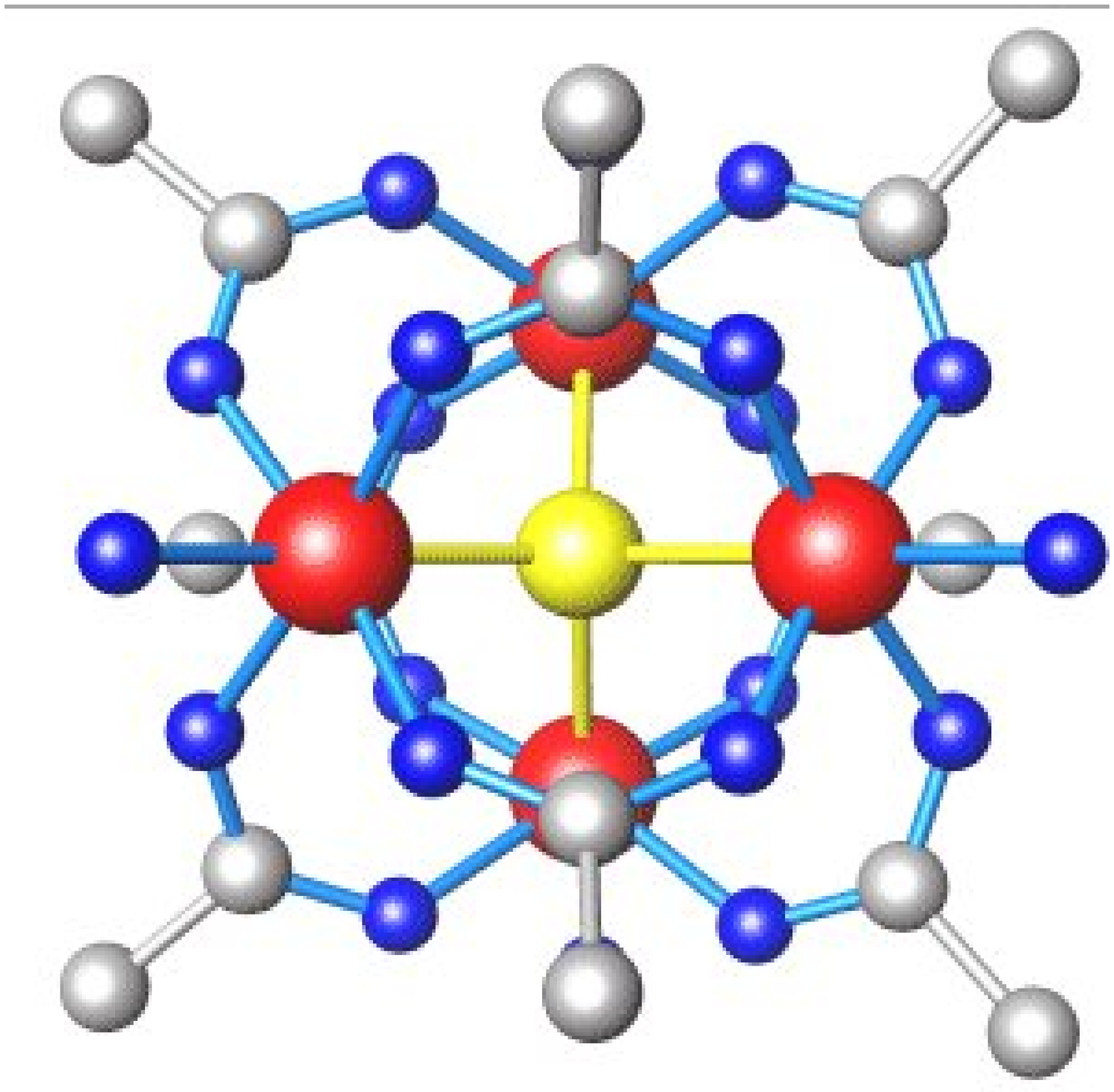}
$\qquad$
\includegraphics[height=45mm]{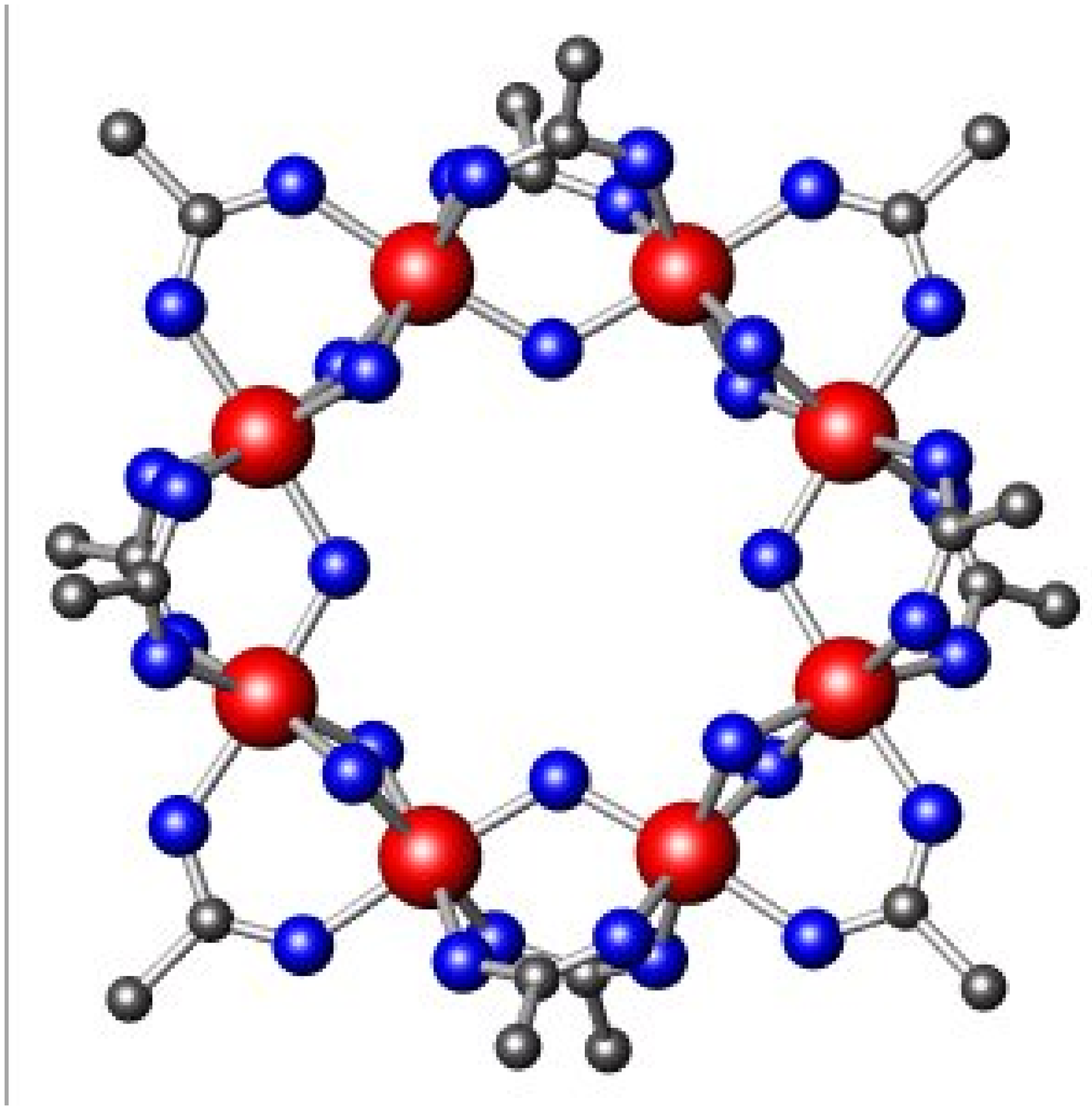}
\caption{Structure of a chromium-4 and a chromium-8 ring. The Cr ions
  are depicted by large spheres.}
\label{magmol-fig:Chromium}
\end{figure}
Another very well investigated class of molecules is given by
spin rings among which iron rings (``ferric wheels") are most
popular
\cite{CTP:JACS93,TDP:JACS94,LGC:PRB97A,LGC:PRB97B,JJL:PRL99,ACC:ICA00,WKS:IO01,MeL:PRL01}.
Iron-6 rings for instance can host alkali ions such as lithium
or sodium which allows to modify the parameters of the spin
Hamiltonian within some range \cite{WSK:IO99,SBC:CEJ01}.
Another realization of rings is possible using chromium ions as
paramagnetic centers. Figure~\ref{magmol-fig:Chromium} shows
the structure of two rings, one with four chromium ions the
other one with eight chromium ions.

\begin{figure}[ht!]
\centering
\includegraphics[height=45mm]{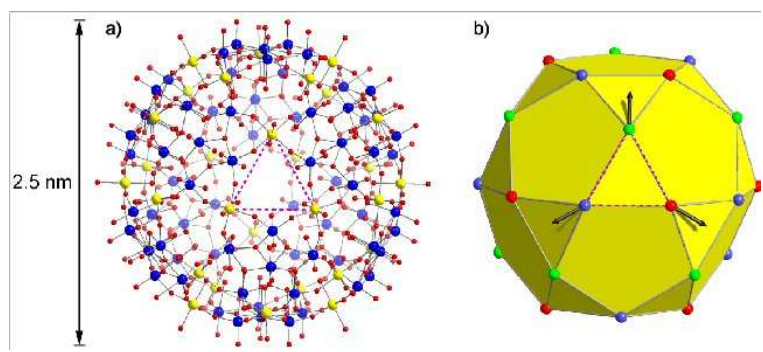}
\caption{Structure of \{Mo$_{72}$Fe$_{30}$\}, a giant Keplerate
  molecule where 30 iron ions are placed at the vertices of an
  icosidodecahedron. L.h.s.: sketch of the chemical structure,
  r.h.s. magnetic structure showing the iron ions (spheres),
  the nearest neighbor interactions (edges) as well as the spin
  directions in the classical ground state. The dashed triangle
  on the l.h.s. corresponds to the respective triangle on the
  r.h.s.. With friendly
  permission by Paul K\"ogerler \cite{MLS:CPC01}.} 
\label{magmol-fig:Fe30}
\end{figure}
A new route to molecular magnetism is based on so-called
Keplerate structures which allow the synthesis of truly giant
and highly symmetric spin arrays. The molecule
\{Mo$_{72}$Fe$_{30}$\} \cite{MSS:ACIE99,MLS:CPC01} containing 30
iron ions of spin $s=5/2$ may be regarded as the archetype of
such structures. Figure~\ref{magmol-fig:Fe30} shows on the
l.h.s. the inner skeleton of this molecule -- Fe and O-Mo-O
bridges -- as well as the classical ground state
\cite{AxL:PRB01} depicted by arrows on the
r.h.s. \cite{MLS:CPC01}.

\begin{figure}[ht!]
\centering
\includegraphics[height=45mm]{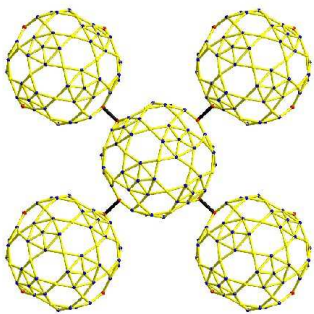}
\caption{Square lattice of \{Mo$_{72}$Fe$_{30}$\}-molecules: Each
  molecule is connected with its four nearest neighbors by an
  antiferromagnetic coupling. With friendly permission by Paul K\"ogerler
  \cite{MSS:ACIE00,MSS:SSS00}.}
\label{magmol-fig:LayeredFe30}
\end{figure}
One of the obvious advantages of magnetic molecules is that the
magnetic centers of different molecules are well separated by
the ligands of the molecules. Therefore, the intermolecular
interactions are utterly negligible and magnetic molecules can
be considered as being independent. Nevertheless, it is
desirable to build up nanostructured materials consisting of
magnetic molecules in a controlled way.
Figure~\ref{magmol-fig:LayeredFe30} gives an example of a planar
structure consisting of layers of \{Mo$_{72}$Fe$_{30}$\}
\cite{MSS:ACIE00,MSS:SSS00} which has been synthesized recently
together with a linear structure consisting of chains of
\{Mo$_{72}$Fe$_{30}$\} \cite{MDT:ACIE02}.  These systems show
new combinations of physical properties that stem from both
molecular and bulk effects.

Many more structures than those sketched above can be
synthesized nowadays and with the increasing success of
coordination chemistry more are yet to come. The final hope of
course is that magnetic structures can be designed according to
the desired magnetic properties. But this goal is not close at
all, it requires further understanding of the interplay of
magneto-chemistry and magnetic phenomena. One of the tools used
to clarify such questions is density functional theory or other
\emph{ab initio} methods
\cite{RAA:JACS97,RCA:JACS98,CSM:CPL99,KHP:PRL01,PBK:PRL02,KBB:PRB02}.

\section{Theoretical techniques and results}
\label{magmol-sec:4}

\subsection{Hamiltonian}
\label{magmol-sec:4-1}

\index{Heisenberg model}

It appears that in the majority of these molecules the
interaction between the localized single-particle magnetic
moments can be rather well described by the Heisenberg model
with isotropic (nearest neighbor) interaction and an additional
anisotropy term
\cite{BeG90,DGP:IC93,CCF:IC95,PDK:JMMM97,WSK:IO99}. Dipolar
interactions are usually of minor importance. It is also found
that antiferromagnetic interactions are favored in most
molecules leading to nontrivial ground states.

\subsubsection*{Heisenberg Hamiltonian}
\label{magmol-sec:4-1-2}

\index{Heisenberg model}

For many magnetic molecules the total Hamilton operator can be
written as
\begin{eqnarray}
\label{magmol-E-4-2}
\op{H}
&=&
\op{H}_{\text{Heisenberg}}
+
\op{H}_{\text{anisotropy}}
+
\op{H}_{\text{Zeeman}}
\\
\label{magmol-E-4-2-H}
\op{H}_{\text{Heisenberg}}
&=&
-
\sum_{u,v}\;
J_{uv}
\op{\vec{s}}(u) \cdot \op{\vec{s}}(v)
\\
\label{magmol-E-4-2-A}
\op{H}_{\text{anisotropy}}
&=&
-
\sum_{u=1}^N\;
d_u (\vec{e}(u)\cdot\op{\vec{s}}(u))^2
\\
\label{magmol-E-4-2-Z}
\op{H}_{\text{Zeeman}}
&=&
g \mu_B \vec{B} \cdot \op{\vec{S}}
\ .
\end{eqnarray}
The Heisenberg Hamilton operator\footnote{Operators are denoted
  by a tilde.} in the form given in 
Eq.~\fmref{magmol-E-4-2-H} is isotropic, $J_{uv}$ is a symmetric
matrix containing the exchange parameters between spins at sites
$u$ and $v$. The exchange parameters are usually given in units of
energy, and $J_{uv}<0$ corresponds to antiferromagnetic,
$J_{uv}>0$ to ferromagnetic coupling\footnote{One has to be
careful with this definition since it varies from author to
author}. The sum in \fmref{magmol-E-4-2-H} runs over all
possible tuples $(u,v)$. The vector operators $\op{\vec{s}}(u)$
are the single-particle spin operators.  

The anisotropy terms \fmref{magmol-E-4-2-A} usually simplify to
a large extend, for instance for spin rings, where the
site-dependent directions $\vec{e}(u)$ are all equal,
e.~g. $\vec{e}(u)=\vec{e}_z$ and the strength as well is the same for all
sites $d_u=d$.

The third part (Zeeman term) in the full Hamiltonian describes
the interaction with the external magnetic field. Without
singe-site and $g$-value anisotropy the direction of the field
can be assumed to be along the $z$-axis which simplifies the
Hamiltonian very much.

Although the Hamiltonian looks rather simple, the eigenvalue
problem is very often not solvable due to the huge dimension of
the Hilbert space or because the number of exchange constants is
too big to allow an accurate determination from experimental
data. Therefore, one falls back to effective single-spin
Hamiltonians for molecules with non-zero ground state spin and a
large enough gap to higher-lying multiplets.

\subsubsection*{Single-spin Hamiltonian}
\label{magmol-sec:4-1-1}

For molecules like Mn$_{12}$ and Fe$_{8}$ which possess a
high ground state spin and well separated higher lying levels
the following single-spin Hamiltonian
\begin{eqnarray}
\label{magmol-E-4-1}
\op{H}
&=&
-D_2 \op{S}_z^2 - D_4 \op{S}_z^4 + \op{H}^\prime
\\
\op{H}^\prime
&=&
g \mu_B B_x \op{S}_x
\end{eqnarray}
is appropriate, see e.~g.~Ref.~\cite{FWK:PRB01}.
The first two terms of the
Hamilton operator
$\op{H}$ represent the anisotropy whereas $\op{H}^\prime$
is the Zeeman term for a magnetic field along the $x$-axis.
The total spin is fixed, i.~e. $S=10$ for Mn$_{12}$ and Fe$_{8}$,
thus the dimension of the Hilbert space is 
$\text{dim}({\mathcal H})=2S+1$. 

The effective Hamiltonian \fmref{magmol-E-4-1} is sufficient to
describe the low-lying spectrum and phenomena like magnetization
tunneling\index{magnetization tunneling}\index{tunneling}. Since
$\op{H}^\prime$ does not commute with the $z$-component of the
total spin $\op{S}_z$, every eigenstate $\ket{M}$ of $\op{S}_z$,
i.~e. the states with good magnetic quantum number $M$, is not
stationary but will tunnel through the barrier and after half
the period be transformed into $\ket{-M}$.

\subsection{Evaluating the spectrum}
\label{magmol-sec:4-2}

The ultimate goal is to evaluate the complete eigenvalue
spectrum of the full Hamilton operator \fmref{magmol-E-4-2} as
well as all eigenvectors. Since the total dimension of the
Hilbert space is usually very large, e.~g. $\text{dim}({\mathcal
  H})=(2s+1)^N$ for a system of $N$ spins of equal spin quantum
number $s$, a straightforward diagonalization of the full
Hamilton matrix is not feasible. Nevertheless, very often the
Hamilton matrix can be decomposed into a block structure because
of spin symmetries or space symmetries.  Accordingly the Hilbert
space can be decomposed into mutually orthogonal subspaces.
Then for a practical evaluation only the size of the largest
matrix to be diagonalized is of importance (relevant dimension).

\subsubsection*{Product basis}
\label{magmol-sec:4-2-0}

The starting point for any diagonalization\index{exact
diagonalization} is the product basis $\ket{\vec{m}}=\ket{m_1,
\dots, m_u, \dots, m_N}$ of the single-particle eigenstates of
all $\op{s}_z(u)$
\begin{eqnarray}
\label{magmol-E-4-5}
\op{s}_z(u)\,
\ket{m_1, \dots, m_u, \dots, m_N}
=
m_u\,
\ket{m_1, \dots, m_u, \dots, m_N}
\ .
\end{eqnarray}
These states are sometimes called Ising states. They span the
full Hilbert space and are used to construct symmetry-related
basis states.

\subsubsection*{Symmetries of the problem}
\label{magmol-sec:4-2-1}

\index{symmetry}
Since the isotropic Heisenberg Hamiltonian \index{Heisenberg model}
includes only a scalar product between spins, this operator is
rotationally invariant in spin space, i.~e. it commutes with
$\op{\vec{S}}$ and thus also with $\op{S}_z$
\begin{eqnarray}
\label{magmol-E-4-3}
\left[\op{H}_{\text{Heisenberg}},\op{\vec{S}}^2\right] 
=
0
\ &,&\qquad
\left[\op{H}_{\text{Heisenberg}},\op{S}_z\right] 
=
0
\ .
\end{eqnarray}
In a case where anisotropy is negligible a well-adapted basis is
thus given by the simultaneous eigenstates $\ket{S,M,\alpha}$ of
$\op{\vec{S}}^2$ and $\op{S}_z$, where $\alpha$ enumerates those
states belonging to the same $S$ and $M$
\cite{Wal:PRB00,BSS:JMMM00}. Since the applied magnetic field
can be assumed to point into $z$-direction for vanishing
anisotropy the Zeeman term automatically also commutes with
$\op{H}_{\text{Heisenberg}}$, $\op{\vec{S}}^2$, and $\op{S}_z$.
Since $M$ is a good quantum number the Zeeman term does not need
to be included in the diagonalization but can be added later.

Besides spin symmetries many molecules possess spatial
symmetries. One example is given by spin rings which have a
translational symmetry. In general the symmetries depend on the
point group of the molecule; for the evaluation of the
eigenvalue spectrum its irreducible representations have to be
used \cite{DGP:IC93,WSK:IO99,Wal:PRB00}. Thus, in a case with
anisotropy one looses spin rotational symmetries but one can
still use space symmetries. Without anisotropy one even gains a
further reduction of the relevant dimension.

\subsubsection*{Dimension of the problem}
\label{magmol-sec:4-2-2}

The following section illuminates the relevant dimensions
assuming certain symmetries\footnote{Work done with Klaus
B\"arwinkel and Heinz-J\"urgen Schmidt, Universit\"at
Osnabr\"uck, Germany.}.

If no symmetry\index{symmetry} is present the total dimension is just
\begin{eqnarray}
\label{magmol-E-4-4}
\text{dim}\left({\mathcal H}\right)
=
\prod_{u=1}^N\;\left(2 s(u) + 1\right)
\end{eqnarray}
for a spin array of $N$ spins with various spin quantum
numbers. In many cases the spin quantum numbers are equal
resulting in a dimension of the total Hilbert space of 
$\text{dim}({\mathcal H})=(2s+1)^N$.

If the Hamiltonian commutes with $\op{S}_z$ then $M$ is a good
quantum number and the Hilbert space ${\mathcal H}$ can be divided
into mutually orthogonal subspaces ${\mathcal H}(M)$
\begin{eqnarray}
\label{magmol-E-4-6}
{\mathcal H}
=
\bigoplus_{M=-\SMax}^{+\SMax}\;
{\mathcal H}(M)
\ ,
\quad
\SMax
=
\sum_{u=1}^N\;s(u)
\ .
\end{eqnarray}
For given values of $M$, $N$ and of all $s(u)$ the dimension
$\text{dim}\left({\mathcal H}(M)\right)$ can be determined as
the number of product states \fmref{magmol-E-4-5}, which
constitute a basis in ${\mathcal H}(M)$, with $\sum_{u} m_u =
M$.  The solution of this combinatorial problem can be given in
closed form \cite{BSS:JMMM00}
\begin{eqnarray}
\label{magmol-E-4-7}
\text{dim}\left({\mathcal H}(M)\right)
=
\frac{1}{(\SMax-M)!}\, 
\left[
\left(\dd{}{z}\right)^{\SMax-M}\;
\prod_{x=1}^{N}\;
\frac{1-z^{2 s(x) + 1}}{1-z}
\right]_{z=0}
\ .
\end{eqnarray}
For equal single-spin quantum numbers
$s(1)=\cdots=s(N)=s$, and thus a maximum total spin quantum
number of $\SMax=Ns$, \eqref{magmol-E-4-7} simplifies to
\begin{eqnarray}
\label{magmol-E-4-8}
\text{dim}\left({\mathcal H}(M)\right)
&=&
f(N,2s+1,\SMax-M)
\qquad\text{with} \\
f(N,\mu,\nu)
&=&
\sum_{n=0}^{\lfloor\nu/\mu\rfloor}
(-1)^n\, \binom{N}{n}
\binom{N-1+\nu - n \mu}{N-1}
\nonumber
\ .
\end{eqnarray}
In both formulae \fmref{magmol-E-4-7} and \fmref{magmol-E-4-8},
$M$ may be replaced by $|M|$ since the dimension of ${\mathcal
H}(M)$ equals those of ${\mathcal
H}(-M)$. $\lfloor\nu/\mu\rfloor$ in the sum symbolizes the
greatest integer less or equal to $\nu/\mu$.
Eq.~\fmref{magmol-E-4-8} is known as a result of de Moivre
\cite{Fel68}.

If the Hamiltonian commutes with $\op{\vec{S}}^2$ and all
individual spins are identical the dimensions of the orthogonal
eigenspaces ${\mathcal H}(S,M)$ can also be determined.  The
simultaneous eigenspaces ${\mathcal H}(S,M)$ of $\op{\vec{S}}^2$
and $\op{S}_z$ are spanned by eigenvectors of $\op{H}$.  The
one-dimensional subspace ${\mathcal H}(M=\SMax)={\mathcal
  H}(\SMax,\SMax)$, especially, is spanned by $\ket{\Omega}$, a
state called magnon vacuum.  The total ladder operators (spin
rising and lowering operators) are
\begin{eqnarray}
\label{magmol-E-4-9}
\op{S}^{\pm}
=
\op{S}_x
\pm
i\,\op{S}_y
\ .
\end{eqnarray}
For $S>M$,  $\op{S}^{-}$ maps any normalized
$\op{H}$-eigenstate 
$\in{\mathcal H}(S,M+1)$ onto an $\op{H}$-eigenstate
$\in{\mathcal H}(S,M)$ with norm $\sqrt{S(S+1)-M(M+1)}$. 

For $0 \le M < \SMax$, ${\mathcal H}(M)$ can be decomposed into
orthogonal subspaces
\begin{eqnarray}
\label{magmol-E-4-10}
{\mathcal H}(M)
=
{\mathcal H}(M,M)
\oplus
\op{S}^{-}{\mathcal H}(M+1)
\end{eqnarray}
with
\begin{eqnarray}
\label{magmol-E-4-11}
\op{S}^{-}{\mathcal H}(M+1)
=
\bigoplus_{S\ge M+1}
{\mathcal H}(S,M)
\ .
\end{eqnarray}
In consequence, the diagonalization\index{exact diagonalization}
of $\op{H}$ in ${\mathcal H}$ has now been traced back to
diagonalization in the subspaces ${\mathcal H}(S,S)$, the
dimension of which are for $S<\SMax$
\begin{eqnarray}
\label{magmol-E-4-12}
\text{dim}\left({\mathcal H}(S,S)\right)
=
\text{dim}\left({\mathcal H}(M=S)\right)
-
\text{dim}\left({\mathcal H}(M=S+1)\right)
\end{eqnarray}
and can be calculated according to \eqref{magmol-E-4-8}.

As an example for space symmetries\index{symmetry} I would like
to discuss the translational symmetry found in spin rings. The
discussed formalism can as well be applied to other symmetry
operations which can be mapped onto a translation. Any such
translation is represented by the cyclic shift operator $\op{T}$
or a multiple repetition. $\op{T}$ is defined by its action on
the product basis \fmref{magmol-E-4-5}
\begin{eqnarray}
\label{magmol-E-4-13}
\op{T}\,
\ket{m_1, \dots, m_{N-1}, m_N}
=
\ket{m_N, m_1, \dots, m_{N-1}}
\ .
\end{eqnarray}
The eigenvalues of $\op{T}$ are the $N$-th roots of unity
\begin{eqnarray}
\label{magmol-E-4-14}
z_k
=
\exp\left\{
-i \frac{2\pi k}{N} 
\right\}
\ ,\qquad
k=0,\dots, N-1
\ ,\qquad
p_k=2\pi k/N
\ ,
\end{eqnarray}
where $k$ will be called translational (or shift) quantum number
and $p_k$ momentum quantum number or crystal momentum.  The
shift operator $\op{T}$ commutes not only with the Hamiltonian
but also with total spin. Any ${\mathcal H}(S,M)$ can therefore
be decomposed into simultaneous eigenspaces ${\mathcal
  H}(S,M,k)$ of $\op{\vec{S}}^2$, $\op{S}_z$ and $\op{T}$.

In the following we demonstrate how an eigenbasis of both
$\op{S}_z$ and $\op{T}$ can be constructed, this basis spans the
orthogonal Hilbert spaces ${\mathcal H}(M,k)$. How total spin
can be included by means of an irreducible tensor operator
approach is described in Refs.~\cite{DGP:IC93,WSK:IO99,Wal:PRB00}.

A special decomposition of ${\mathcal H}$ into orthogonal
subspaces can be achieved by starting with the product basis and
considering the equivalence relation
\begin{eqnarray}
\label{magmol-E-4-99}
\ket{\psi}\cong\ket{\phi}
\Leftrightarrow
\ket{\psi}=\op{T}^n\,\ket{\phi}
\ ,\ n\in\{ 1,2,\dots,N\}
\end{eqnarray}
for any pair of states belonging to the product basis. The
equivalence relation then induces a complete decomposition of
the basis into disjoint subsets, i.~e. the equivalence classes.
A ``cycle" is defined as the linear span of such an equivalence
class of basis vectors. The obviously orthogonal decomposition
of ${\mathcal H}$ into cycles is compatible with the
decomposition of ${\mathcal H}$ into the various ${\mathcal
H}(M)$. Evidently, the dimension of a cycle can never exceed
$N$. Cycles are called ``proper cycles" if their dimension
equals $N$, they are termed ``epicycles" else. One of the $N$
primary basis states of a proper cycle may arbitrarily be
denoted as
\begin{eqnarray}
\label{magmol-E-4-15}
\ket{\psi_1}
=
\ket{m_1, \dots, m_N}
\end{eqnarray}
and the remaining ones may be enumerated as 
\begin{eqnarray}
\label{magmol-E-4-16}
\ket{\psi_{n+1}}
=
\op{T}^n\,\ket{\psi_1}
\ ,\ n=1,2,\dots,N-1
\ .
\end{eqnarray}
The cycle under consideration is likewise spanned by the states 
\begin{eqnarray}
\label{magmol-E-4-17}
\ket{\chi_k}
=
\frac{1}{\sqrt{N}}
\sum_{\nu=0}^{N-1}\,
\left(
e^{i\frac{2\pi\,k}{N}}
\op{T}
\right)^{\nu}
\ket{\psi_{1}}
\end{eqnarray}
which are eigenstates of $\op{T}$ with the respective shift
quantum number $k$. Consequently, every $k$ occurs once in a
proper cycle. An epicycle of dimension $D$ is spanned by $D$
eigenstates of $\op{T}$ with each of the translational quantum
numbers $k=0,N/D,\dots,(D-1)N/D$ occurring exactly once.

As a rule of thumb one can say that the dimension of each
${\mathcal H}(M,k)$ is approximately $\text{dim}({\mathcal
H}(M,k))\approx \text{dim}({\mathcal H}(M))/N$. An exact
evaluation of the relevant dimensions for spin rings can be
obtained from Ref.~\cite{BSS:JMMM00}.

\subsubsection*{Exact diagonalization}
\label{magmol-sec:4-2-3}

\index{exact diagonalization}

If the relevant dimension is small enough the respective
Hamilton matrices can be diagonalized, either analytically
\cite{Kou:JMMM97,Kou:JMMM98,BSS:JMMM00} or numerically, see
e.~g.
\cite{BoF:PR64,BoJ:PRB83,FLM:PRB91,Man:RMP91,DGP:IC93,GJL:PRB94,FLS:PRB97,Wal:PRB00}.

Again, how such a project is carried out, will be explained with
the help of an example, a simple spin ring with $N=6$ and
$s=1/2$. The total dimension is $\text{dim}({\mathcal
H})=(2s+1)^N=64$. The Hamilton operator \fmref{magmol-E-4-2-H}
\index{Heisenberg model} simplifies to
\begin{eqnarray}
\label{magmol-E-4-18}
\op{H}_{\text{Heisenberg}}
&=&
- 2 J
\sum_{u=1}^N\;
\op{\vec{s}}(u) \cdot \op{\vec{s}}(u+1)
\ ,\quad
N+1\equiv 1
\ .
\end{eqnarray}
We start with the magnon vacuum $\ket{\Omega}=\ket{++++++}$
which spans the Hilbert space ${\mathcal H}(M)$ with $M=Ns=3$.
``$\pm$" are shorthand notations for $m=\pm 1/2$. The dimension
of the subspace $\text{dim}({\mathcal H}(M=Ns))$ is one and the
energy eigenvalue is $E_{\Omega}=-2JNs^2=-3J$. $\ket{\Omega}$ is
an eigenstate of the shift operator with $k=0$. Since $S$ is
also a good quantum number in this example $\ket{\Omega}$ has to
be an eigenstate of $\op{\vec{S}}^2$, too, the quantum number is
$S=Ns$.

The next subspace ${\mathcal H}(M)$ with $M=Ns-1=2$ is spanned
by $\ket{-+++++}$ and the five other vectors which are obtained
by repetitive application of $\op{T}$. This subspace obviously
has the dimension $N$, and the cycle spanned by
$\op{T}^n\ket{-+++++}, n=0,\dots,N-1$ is a proper
one. Therefore, each $k$ quantum number arises once. The
respective eigenstates of $\op{T}$ can be constructed according
to Eq.~\fmref{magmol-E-4-17} as
\begin{eqnarray}
\label{magmol-E-4-19}
\ket{M=2,k}
&=&
\frac{1}{\sqrt{N}}
\sum_{\nu=0}^{N-1}\,
\left(
e^{i\frac{2\pi\,k}{N}}
\op{T}
\right)^{\nu}
\ket{-+++++}
\ .
\end{eqnarray}
All subspaces ${\mathcal H}(M,k)$ have dimension one. Since
$\op{S}^-\ket{\Omega}$ is a state belonging to ${\mathcal
H}(M=Ns-1)$ with the same $k$-quantum number as $\ket{\Omega}$
it is clear that $\ket{M=2,k=0}$ is a already an eigenstate of
$\op{\vec{S}}^2$ with $S=Ns$. The other $\ket{M=2,k\ne0}$ must
have $S=Ns-1$.

The next subspace ${\mathcal H}(M)$ with $M=Ns-2=1$ is spanned
by three basic vectors, i.~e. $\ket{--++++}, \ket{-+-+++},
\ket{-++-++}$ and the repetitive applications of $\op{T}$ onto
them. The first two result in proper cycles, the third vector
$\ket{-++-++}$ results in an epicycle of dimension three, thus
for the epicycle we find only $k$ quantum numbers $k=0,2,4$. The
energy eigenvalues found in the subspace ${\mathcal H}(M=Ns-1)$
(``above") must reappear here which again allows to address an
$S$ quantum number to these eigenvalues. The dimension of the
subspace ${\mathcal H}(M=1)$ is 15, the dimensions of the
subspaces ${\mathcal H}(M,k)$ are 3 ($k=0$), 2 ($k=1$), 3
($k=2$), 2 ($k=3$), 3 ($k=4$), and 2 ($k=5$).

The last subspace which has to be considered belongs to $M=0$
and is spanned by $\ket{---+++}, \ket{--+-++}, \ket{-+--++},
\ket{-+-+-+}$ and repetitive applications of $\op{T}$. Its
dimension is 20. Here $\ket{-+-+-+}$ leads to an epicycle of
dimension two. 

The Hamilton matrices in subspaces with $M<0$ need not to be
diagonalized due to the $\op{S}_z$-symmetry\index{symmetry},
i.~e. eigenstates with negative $M$ can be obtained by
transforming all individual $m_u \rightarrow -m_u$. 
Summing up the dimensions of all ${\mathcal H}(M)$ then yields 
$1+6+15+20+15+6+1=64\;\surd$.

\begin{figure}[ht!]
\centering
\includegraphics[height=35mm]{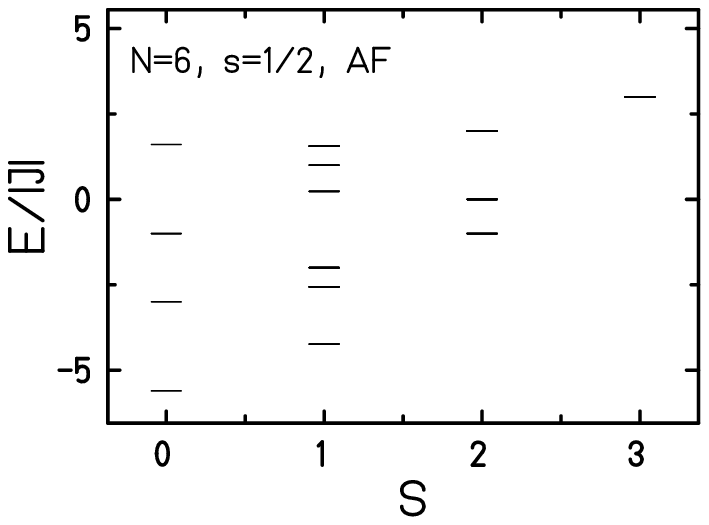}
$\qquad$
\includegraphics[height=35mm]{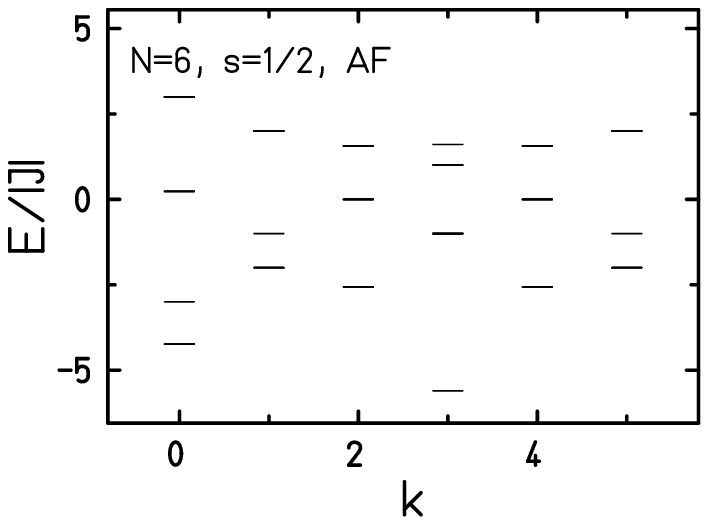}
\caption{Energy eigenvalues as a function of total spin quantum
  number $S$ (l.h.s.) and $k$ (r.h.s.).}
\label{magmol-fig:Ring-6-1}
\end{figure}
Figure~\ref{magmol-fig:Ring-6-1} shows the resulting energy
spectrum both as a function of total spin $S$ as well as a
function of translational quantum number $k$.

\subsubsection*{Projection and \Lanczos\ method}
\label{magmol-sec:4-2-4}

\index{Lanczos} Complex hermitian matrices can be completely
diagonalized\index{exact diagonalization} numerically up to a
size of about 10,000 by 10,000 which corresponds to about
1.5~Gigabyte of necessary RAM. Nevertheless, for larger systems
one can still use numerical methods to evaluate low-lying energy
levels and the respective eigenstates with high accuracy.

A simple method is the projection method \cite{Man:RMP91} which
rests on the multiple application of the Hamiltonian on some
random trial state.

To be more specific let's approximate the ground state of a spin
system. We start with a random trial state $\ket{\phi_0}$ and
apply an operator which ``cools" the system. This operator is
given by the time evolution operator with imaginary time steps
\begin{eqnarray}
\label{magmol-E-4-20}
\ket{\tilde{\phi}_1}
&=&
\exp\left\{-\varepsilon \op{H}\right\}
\ket{\phi_0}
\ .
\end{eqnarray}
Expanding $\ket{\phi_0}$ into eigenstates $\ket{\nu}$ of the Hamilton
operator elucidates how the method works
\begin{eqnarray}
\label{magmol-E-4-21}
\ket{\tilde{\phi}_1}
&=&
\sum_{\nu=0}
\exp\left\{-\varepsilon E_{\nu}\right\}
\ket{\nu}
\braket{\nu}{\phi_0}
\\
&=&
\exp\left\{-\varepsilon E_{0}\right\}
\sum_{\nu=0}
\exp\left\{-\varepsilon (E_{\nu}-E_{0})\right\}
\ket{\nu}
\braket{\nu}{\phi_0}
\ .
\end{eqnarray}
Relative to the ground state component all other components are
exponentially suppressed. For practical purposes equation
\fmref{magmol-E-4-21} is linearized and recursively used
\begin{eqnarray}
\label{magmol-E-4-22}
\ket{\tilde{\phi}_{i+1}}
&=&
\left(
1-\varepsilon \op{H}
\right)
\ket{\phi_i}
\ ,\quad
\ket{\phi_{i+1}}
=
\frac{\ket{\tilde{\phi}_{i+1}}}{\sqrt{\braket{\tilde{\phi}_{i+1}}{\tilde{\phi}_{i+1}}}}
\ .
\end{eqnarray}
$\varepsilon$ has to be small enough in order to allow the
linearization of the exponential. It is no problem to evaluate
several higher-lying states by demanding that they have to be
orthogonal to the previous ones. Restricting the calculation to
orthogonal eigenspaces yields low-lying states in these
eigenspaces which allows to evaluate even more energy levels.
The resulting states obey the properties of the Ritz variational
principle, i.~e. they lie above the ground state and below the
highest one.

\index{Lanczos} Another method to partially
diagonalize\index{exact diagonalization} a huge matrix was
proposed by Cornelius \Lanczos\ in 1950
\cite{Lan:JRNBS50,CuW85}. Also this method uses a (random)
initial vector. It then generates an orthonormal system in such
a way that the representation of the operator of interest is
tridiagonal. Every iteration produces a new tridiagonal matrix
which is by one row and one column bigger than the previous
one. With growing size of the matrix its eigenvalues converge
against the true ones until, in the case of finite dimensional
Hilbert spaces, the eigenvalues reach their true values. The key
point is that the extremal eigenvalues converge rather quickly
compared to the other ones \cite{BDD00}. Thus it might be that
after 300 \Lanczos\ steps the ground state energy is already
approximated to 10 figures although the dimension of the
underlying Hilbert space is $10^8$.

A simple \Lanczos\ algorithm looks like the following. One
starts with an arbitrary vector $\ket{\psi_0}$, which has to
have an overlap with the (unknown) ground state.
The next orthogonal vector is constructed by application of 
$\op{H}$ and projecting out the original vector $\ket{\psi_0}$
\begin{equation}
\label{magmol-E-4-24}
\ket{\psi_1'}
=
\left(1-\ket{\psi_0}\bra{\psi_0}\right)\op{H}\ket{\psi_0}
=
\op{H}\ket{\psi_0}-\bra{\psi_0}\op{H}\ket{\psi_0}\ket{\psi_0}
\ ,
\end{equation}
which yields the normalized vector
\begin{equation}
\label{magmol-E-4-25}
\ket{\psi_{1}}
=
\frac{\ket{\psi_{1}'}}{\sqrt{\braket{\psi_{1}'}{\psi_{1}'}}}
\ .
\end{equation}
Similarly all further basis vectors are generated
\begin{eqnarray}
\label{magmol-E-4-26}
\ket{\psi_{k+1}'} 
& = & \left(1-\ket{\psi_k}\bra{\psi_k}
-\ket{\psi_{k-1}}\bra{\psi_{k-1}} \right)\op{H}\ket{\psi_k} 
\\
& = & \op{H}\ket{\psi_k}-\bra{\psi_k}H\ket{\psi_k}\ket{\psi_k}
-\bra{\psi_{k-1}}\op{H}\ket{\psi_k}\ket{\psi_{k-1}}
\nonumber
\end{eqnarray}
and
\begin{equation}
\label{magmol-E-4-27}
\ket{\psi_{k+1}}=\frac{\ket{\psi_{k+1}'}}{\sqrt{\braket{\psi_{k+1}'}
{\psi_{k+1}'}}}
\ .
\end{equation}
The new \Lanczos\ vector is by construction orthogonal to the two
previous ones. Without proof we repeat that it is then also
orthogonal to all other previous \Lanczos\ vectors. This
constitutes the tridiagonal form of the resulting Hamilton
matrix
\begin{equation}
\label{magmol-E-4-28}
T_{i,j}=\bra{\psi_i}\op{H}\ket{\psi_j} \quad\quad {\textrm{with}} \quad\quad T_{i,j}=0
\quad {\textrm{if}} \quad \left|i-j\right|>1
\ .
\end{equation}
The \Lanczos\ matrix $T$ can be diagonalized at any
step. Usually one iterates the method until a certain
convergence criterion is fulfilled.

The eigenvectors of $\op{H}$ can be approximated using the
eigenvectors $\ket{\phi_\mu}$ of $T$
\begin{equation}
\label{magmol-E-4-29}
\ket{\chi_\mu}\approx\sum_{i=0}^n{\braket{\psi_i}{\phi_\mu}\ket{\psi_i}}
\ ,
\end{equation}
where $\mu$ labels the desired energy eigenvalue, e.~g. the
ground state energy. $n$ denotes the number of iterations.

The simple \Lanczos\ algorithm has some problems due to limited
accuracy. One problem is that eigenvalues may collapse. Such
problems can be solved with more refined formulations of the
method \cite{CuW85}.

\subsubsection*{DMRG}
\label{magmol-sec:4-2-5}

\index{DMRG}

The DMRG technique \cite{Whi:PRB93} has become one of the
standard numerical methods for quantum lattice calculations in
recent years \cite{PWK99,Sch:RMP04}. Its basic idea is the reduction of
Hilbert space while focusing on the accuracy of a target state.
For this purpose the system is divided into subunits -- blocks
-- which are represented by reduced sets of basis states. The
dimension $m$ of the truncated block Hilbert space is a major
input parameter of the method and to a large extent determines
its accuracy.
 
DMRG is best suited for chain-like structures. Many accurate
results have been achieved by applying DMRG to various
(quasi-)one-dimensional systems
\cite{WhD:PRB93B,GJL:PRB94,Xia:PRB98}. The best results were
found for the limit of infinite chains with open boundary
conditions. It is commonly accepted that DMRG reaches maximum
accuracy when it is applied to systems with a small number of
interactions between the blocks, e.~g. systems with only
nearest-neighbor interaction \cite{PWK99}.

\begin{figure}[ht!]
\begin{center}
\scalebox{0.23}{
\includegraphics[clip]{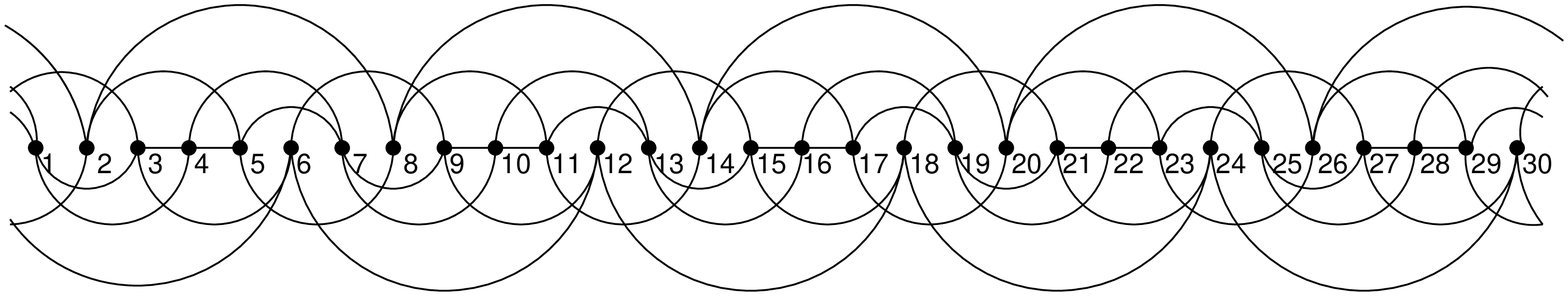}
}
\end{center}
\caption{\label{magmol-fig:fe_30_projection_1d.eps}One-dimensional projection of the
icosidodecahedron: the lines represent interactions.}
\end{figure}

It is not \emph{a priori} clear how good results for finite systems
like magnetic molecules are\footnote{Work done with Matthias
Exler, Universit\"at Osnabr\"uck, Germany.}. Such systems are
usually not chain-like, so in order to carry out DMRG
calculations a mapping onto a one-dimensional structure has to
be performed \cite{PWK99}.  Since the spin array consists of a
countable number of spins, any arbitrary numbering is already a
mapping onto a one-dimensional structure. However, even if the
original system had only nearest-neighbor exchange, the new
one-dimensional system has many long-range interactions
depending on the way the spins are enumerated. Therefore, a
numbering which minimizes long range interactions is
preferable. Fig.~\ref{magmol-fig:fe_30_projection_1d.eps} shows
the graph of interactions for the molecule
\{Mo$_{72}$Fe$_{30}$\} which we want to consider as an example
in the following \cite{ExS:PRB03}.

\begin{figure}[ht!]
\begin{center}
\scalebox{0.5}{
\includegraphics[clip]{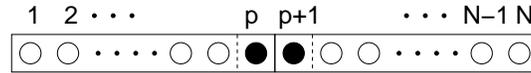}
}
\end{center}
\caption{\label{magmol-fig:block_algo_2d.eps}Block setup for DMRG ``sweep'' algorithm:
  The whole system of $N$ spins constitutes the
  superblock. The spins $\{1,2,\ldots,p\}$ belong to the left block,
  the other spins $\{p+1,\ldots,N\}$ to the right block.}
\end{figure}

For finite systems a block algorithm including sweeps, which is
similar to the setup in White's original article
\cite{Whi:PRB93}, has turned out to be most efficient. Two
blocks are connected via two single spin sites, these four parts
form the superblock see Fig.~\ref{magmol-fig:block_algo_2d.eps}.

For illustrative purposes we use a simple Heisenberg
Hamiltonian\index{Heisenberg model}, compare
\fmref{magmol-E-4-2-H}. The Hamiltonian is invariant under
rotations in spin space.  Therefore, the total magnetic quantum
number $M$ is a good quantum number and we can perform our
calculation in each orthogonal subspace ${\mathcal H}(M)$
separately.

Since it is difficult to predict the accuracy of a DMRG
calculation, it is applied to an exactly diagonalizable system
first. The most realistic test system for the use of DMRG for
\mofe\ is the icosidodecahedron with spins $s=1/2$. This
fictitious molecule, which possibly may be synthesized with
vanadium ions instead of iron ions, has the same structure as
\mofe, but the smaller spin quantum number reduces the dimension
of the Hilbert space significantly. Therefore a numerically
exact determination of low-lying levels using a
\Lanczos\ \index{Lanczos} method is possible
\cite{SSR:EPJB01}. These results are used to analyze the
principle feasibility and the accuracy of the method.

\begin{figure}
\begin{center}
\scalebox{0.18}{
\includegraphics[clip]{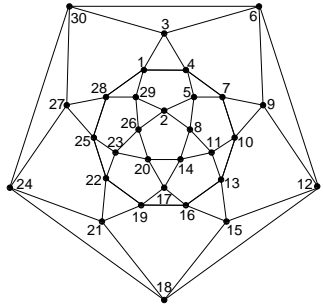}
}
\end{center}
\caption{\label{magmol-fig:icosi-graph.eps}Two-dimensional
projection of the icosidodecahedron, the site numbers are those
used in our DMRG algorithm.}
\end{figure}
The DMRG calculations were implemented using the enumeration of
the spin sites as shown in
Figs.~\ref{magmol-fig:fe_30_projection_1d.eps}
and~\ref{magmol-fig:icosi-graph.eps}.  This enumeration
minimizes the average interaction length between two sites. 

\begin{figure}[ht!]
\begin{center}
\scalebox{0.3}{
\includegraphics[clip]{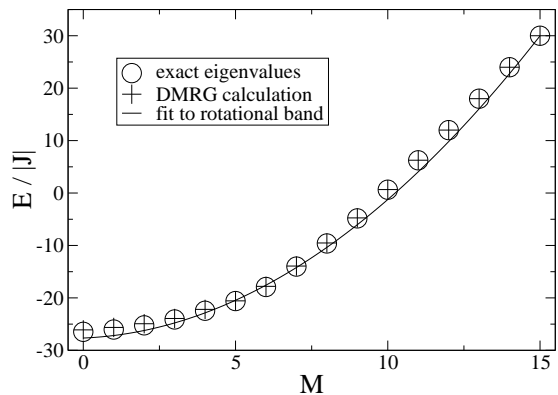}
}
\end{center}
\caption{\label{magmol-fig:icosi_S_0_5_m_60_rot_band.eps}
Minimal energy eigenvalues of the $s=1/2$ icosidodecahedron. The DMRG
result with $m=60$ is depicted by crosses, the \Lanczos\ values
by circles. The rotational band is discussed in subsection
\ref{magmol-sec:4-3-2}.}
\end{figure}
\begin{figure}[ht!]
  \begin{center}
    \scalebox{0.3}{
      \includegraphics[clip]{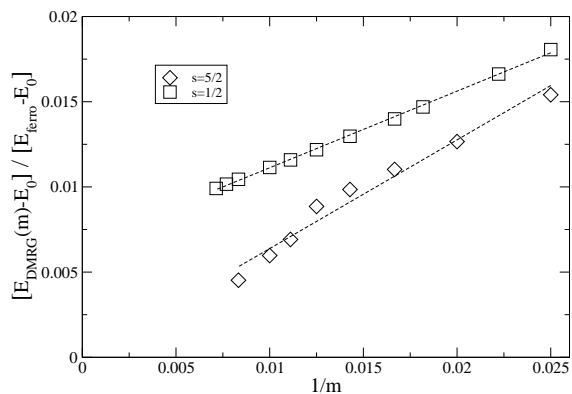}
      }
    \caption{\label{magmol-fig:m_extrapol_both_s.eps}Dependence of the
      approximate ground state energy on the DMRG parameter
      $m$. $E_0$ is the true ground state energy in the case
      $s=1/2$ and the extrapolated one for $s=5/2$.}
  \end{center}
\end{figure}

In Fig.~\ref{magmol-fig:icosi_S_0_5_m_60_rot_band.eps} the DMRG
results (crosses) are compared to the energy eigenvalues
(circles) determined numerically with a \Lanczos\ method
\cite{SSR:EPJB01,ExS:PRB03}. Very good agreement of both
sequences, with a maximal relative error of less than 1\% is
found. Although the high accuracy of one-dimensional
calculations (often with a relative error of the order of
$10^{-6}$) is not achieved, the result demonstrates that DMRG is
applicable to finite 2D spin systems.  Unfortunately, increasing
$m$ yields only a weak convergence of the relative error, which
is defined relative to the width of the spectrum
\begin{equation}
\label{magmol-E-4-23}
\epsilon\left(m\right)=\frac{E_{\rm{DMRG}}\left(m\right)-E_0}
{\left|E_0^{\rm{AF}}-E_0^{\rm{F}}\right|}
\ .
\end{equation}
The dependence for a quasi two-dimensional structure like the
icosidodecahedron is approximately proportional to $1/m$ (see
Fig.~\ref{magmol-fig:m_extrapol_both_s.eps}).  Unfortunately,
such weak convergence is characteristic for two-dimensional
systems in contrast to one-dimensional chain structures, where
the relative error of the approximate energy was reported to
decay exponentially with $m$ \cite{Whi:PRB93}. Nevertheless, the
extrapolated ground state energy for $s=1/2$ deviates only by
$\epsilon=0.7$~\% from the ground state energy determined with a
\Lanczos\ algorithm.

The major result of the presented investigation is that the DMRG
approach delivers acceptable results for finite systems like
magnetic molecules. Nevertheless, the accuracy known from
one-dimensional systems is not reached.

\subsubsection*{Spin-coherent states}
\label{magmol-sec:4-2-6}

Spin-coherent states \cite{Rad:JPA71} provide another means to
either treat a spin system exactly and investigate for instance
its dynamics \cite{HML:EPJB02} or to use spin coherent states in
order to approximate the low-lying part of the spectrum. They
are also used in connection with path integral methods.  In the
following the basic ideas and formulae will be presented.

The obvious advantage of spin-coherent states is that they
provide a bridge between classical spin dynamics and quantum
spin dynamics. Spin coherent states are very intuitive since
they parameterize a quantum state by the expectation value of
the spin operator, e.~g. by the two angles which represent the
spin direction.

Spin coherent states $\ket{z}$ are defined as
\begin{eqnarray}
\label{magmol-E-4-30}
\ket{z} 
&=& 
\frac{1}{(1+|z|^{2})^{s}} 
\sum_{p=0}^{2s} \sqrt{\binom{2s}{p}}~z^{p}~\ket{s,m=s-p}
\ ,\quad
z \in \mathbb{C}
\ .
\end{eqnarray}
In this definition spin-coherent states are characterized by the
spin length $s$ and a complex value $z$. The states
\fmref{magmol-E-4-30} are normalized but not orthogonal
\begin{eqnarray}
\label{magmol-E-4-31}
\braket{z}{z}
&=&
1
\ ,\quad
\braket{y}{z}
=
\frac{(1+y^{*}z)^{2s}}{(1+|y|^{2})^{s}
(1+|z|^{2})^{s}} 
\nonumber
\ .
\end{eqnarray}
Spin-coherent states provide a basis in single-spin Hilbert
space, but they form an overcomplete set of states. Their
completeness relation reads
\begin{eqnarray}
\label{magmol-E-4-32}
\EinsOp 
&=&
\frac{2s+1}{\pi}
\int
\dint^{2}z
\frac{\ket{z}\bra{z}}{(1+|z|^{2})^{2}}
\ ,\quad
\dint^{2}z = \dint\mathrm{Re}(z)~\dint\mathrm{Im}(z)
\ .
\end{eqnarray}
The intuitive picture of spin-coherent states becomes obvious if
one transforms the complex number $z$ into angles on a Riemann
sphere 
\begin{eqnarray}
\label{magmol-E-4-33}
z = \tan(\theta/2)e^{i \phi}
\ ,\quad
0 \leq \theta < \pi 
\ , \quad
0 \leq \phi < 2\pi
\ .
\end{eqnarray}
Thus, spin-coherent states may equally well be represented by
two polar angles $\theta$ and $\phi$.
Then the expectation value of the spin operator $\op{\vec{s}}$
is simply
\begin{eqnarray}
\label{magmol-E-4-34}
\bra{\theta, \phi}
\op{\vec{s}}
\ket{\theta, \phi}
&=&
s
\begin{pmatrix}
\sin(\theta) \cos(\phi)\\
\sin(\theta) \sin(\phi)\\
\cos(\theta)
\end{pmatrix}
\ .
\end{eqnarray}
Using \fmref{magmol-E-4-33} the definition of the states
$\ket{\theta, \phi}$ which is equivalent to
Eq.~\fmref{magmol-E-4-30} is then given by
\begin{eqnarray}
\label{magmol-E-4-35}
\ket{\theta, \phi}
&=&
\sum_{p=0}^{2s}\sqrt{\binom{2s}{p}}
\left[\cos(\theta/2)\right]^{(2s-p)}
\left[e^{i \phi}\sin(\theta/2)\right]^{p}\ket{s, m=s-p}
\end{eqnarray}
and the completeness relation simplifies to
\begin{eqnarray}
\label{magmol-E-4-36}
\EinsOp
&=&
\frac{2s+1}{4 \pi}
\int\dint\Omega~\ket{\theta, \phi}\bra{\theta, \phi}
\ .
\end{eqnarray}
Product states of spin-coherent states span the many-spin
Hilbert space. A classical ground state can easily be translated
into a many-body spin-coherent state. One may hope that this
state together with other product states can provide a useful
set of linearly independent states in order to approximate
low-lying states of systems which are too big to handle
otherwise. But it is too early to judge the quality of such
approximations.

\subsection{Evaluation of thermodynamic observables}
\label{magmol-sec:4-5}

For the sake of completeness we want to outline how basic
observables can be evaluated both as function of temperature $T$
and magnetic field $B$. We will assume that
$\left[\op{H},\op{S}_z\right]=0$ for this part, so that the
energy eigenvectors $\ket{\nu}$ can be chosen as simultaneous
eigenvectors of $\op{S}_z$ with eigenvalues $E_{\nu}(B)$ and
$M_{\nu}$. The energy dependence of $E_{\nu}(B)$ on $B$ is
simply given by the Zeeman term. If $\op{H}$ and $\op{S}_z$ do
not commute the respective traces for the partition function and
thermodynamic means have to be evaluated starting from their
general definitions.

The partition function is defined as follows
\begin{eqnarray}
\label{magmol-E-4-100}
Z(T,B)
&=&
\tr\left\{\mbox{e}^{-\beta\op{H}}\right\}
=
\sum_{\nu}\,\mbox{e}^{-\beta E_{\nu}(B)}
\ .
\end{eqnarray}

Then the magnetization and the susceptibility per molecule
can be evaluated from the first and the second moment of
$\op{S}_z$ 
\begin{eqnarray}
\label{magmol-E-4-101}
{\mathcal M}(T,B)
&=&
-
\frac{1}{Z}\,
\tr\left\{g \mu_B \op{S}_z \mbox{e}^{-\beta\op{H}}\right\}
\\
\nonumber
&=&
-\frac{g \mu_B}{Z}\,
\sum_{\nu}\,M_{\nu}\, \mbox{e}^{-\beta E_{\nu}(B)}
\\
\label{magmol-E-4-102}
\chi(T,B)
&=&
\frac{\partial {\mathcal M}(T,B)}{\partial B}
\\
\nonumber
&=&
\frac{(g \mu_B)^2}{k_B T}\,
\Bigg\{
\frac{1}{Z}\,
\sum_{\nu}\,M_{\nu}^2\, \mbox{e}^{-\beta E_{\nu}(B)}
-
\left(
\frac{1}{Z}\,
\sum_{\nu}\,M_{\nu}\, \mbox{e}^{-\beta E_{\nu}(B)}
\right)^2
\Bigg\}
\ .
\end{eqnarray}
In a similar way the internal energy and the specific heat are
evaluated from first and second moment of the Hamiltonian
\begin{eqnarray}
\label{magmol-E-4-103}
U(T,B)
&=&
-
\frac{1}{Z}\,
\tr\left\{\op{H} \mbox{e}^{-\beta\op{H}}\right\}
\\
\nonumber
&=&
-\frac{1}{Z}\,
\sum_{\nu}\,E_{\nu}(B)\, \mbox{e}^{-\beta E_{\nu}(B)}
\\
\label{magmol-E-4-104}
C(T,B)
&=&
\frac{\partial U(T,B)}{\partial B}
\\
\nonumber
&=&
\frac{1}{k_B T^2}\,
\Bigg\{
\frac{1}{Z}\,
\sum_{\nu}\,\left(E_{\nu}(B)\right)^2\, \mbox{e}^{-\beta E_{\nu}(B)}
-
\left(
\frac{1}{Z}\,
\sum_{\nu}\,E_{\nu}(B)\, \mbox{e}^{-\beta E_{\nu}(B)}
\right)^2
\Bigg\}
\ .
\end{eqnarray}

\subsection{Properties of spectra}
\label{magmol-sec:4-3}

In the following chapter I am discussing some properties of the
spectra of magnetic molecules with isotropic and
antiferromagnetic interaction.
\index{Heisenberg model}

\subsubsection*{Non-bipartite spin rings}
\label{magmol-sec:4-3-1}

With the advent of magnetic molecules it appears to be possible
to synthesize spin rings with an odd number of spins. Although
related to infinite spin rings and chains such systems have not
been considered mainly since it does not really matter whether
an infinite ring has an odd or an even number of spins. In
addition the sign rule of Marshall and Peierls \cite{Mar:PRS55}
and the famous theorems of Lieb, Schultz, and Mattis
\cite{LSM:AP61,LiM:JMP62} provided valuable tools for the
understanding of even rings which have the property to be
bipartite and are thus non-frustrated. These theorems explain
the degeneracy of the ground states in subspaces ${\mathcal
  H}(M)$ as well as their shift quantum number $k$ or
equivalently crystal momentum quantum number $p_k=2\pi k/N$.

\begin{table}[ht!]
\begin{center}
\begin{tabular}{|cc||r|r|r|r|r|r|r|r|r|l|}
\hline
&$s$&\multicolumn{9}{c|}{$N$}&\\
&& 2&3&4&5&6&7&8&9&10&\\
\hline\hline
\multirow{36}{0mm}{}
&             & 1.5 & 0.5  & 1 & 0.747 & 0.934 & 0.816 
& 0.913 & 0.844 & 0.903 & $E_0/(NJ)$\\
&$\frac{1}{2}$& 1   & 4    & 1 & 4        & 1        & 4        
& 1        & 4        & 1 & deg\\
&             & 0 &1/2& 0 & 1/2 & 0   & 1/2 & 0   & 1/2 & 0   & $S$\\
&             & 1   & 1, 2 & 0 & 1, 4     & 3        & 2, 5
& 0        & 2, 7     & 5 & $k$\\
\cline{2-12}
&             &4.0&3.0&2.0&2.236&1.369&2.098&1.045&1.722&0.846& $\Delta E/|J|$\\
&$\frac{1}{2}$& 3 & 4 & 3 &2    & 3   & 8   & 3   & 8   & 3   & deg\\
&             & 1 &3/2& 1 & 1/2 & 1   & 3/2 & 1   & 3/2 & 1   & $S$\\
&             & 0 & 0 & 2 & 0   & 0   &1, 6 & 4   & 3, 6& 0   & $k$\\
\cline{1-12}
&   & 4 & 2 & 3 & 2.612 &2.872&2.735&2.834&2.773&2.819& $E_0/(NJ)$\\
&$1$& 1 & 1 & 1 & 1     & 1   & 1   & 1   & 1   & 1   &  deg\\
&   & 0 & 0 & 0 & 0     & 0   & 0   & 0   & 0   & 0   & $S$\\
&   & 0 & 0 & 0 & 0     & 0   & 0   & 0   & 0   & 0   & $k$\\
\cline{2-12}
&   &4.0&2.0    &2.0&1.929&1.441&1.714&1.187&1.540&1.050& $\Delta E/|J|$\\
&$1$& 3 & 9     & 3 & 6   & 3   & 6   & 3   & 6   & 3   &  deg\\
&   & 1 & 1     & 1 & 1   & 1   & 1   & 1   & 1   & 1   & $S$\\
&   & 1 &0, 1, 2& 2 & 2, 3& 3   &3, 4 & 4   &4, 5 & 5   & $k$\\
\hline
\end{tabular}
\vspace*{5mm}
\end{center}
\caption[]{Properties of ground and first excited state of AF
Heisenberg rings for various $N$ and $s$: ground state energy
$E_0$, gap $\Delta E$, degeneracy $deg$, total spin $S$ and
shift quantum number $k$.
}\label{magmol:T-2-1-A}
\end{table}

\begin{table}[ht!]
\begin{center}
\begin{tabular}{|cc||r|r|r|r|r|r|r|r|r|l|}
\hline
&$s$&\multicolumn{9}{c|}{$N$}&\\
&& 2&3&4&5&6&7&8&9&10&\\
\hline\hline
\multirow{36}{0mm}{}
&             & 7.5 & 3.5 & 6 &4.973&5.798&5.338&5.732&5.477&$5.704^{\dagger\dagger}$& $E_0/(NJ)$\\
&$\frac{3}{2}$& 1   & 4   & 1 & 4   & 1   & 4   & 1   & 4   & 1 &  deg\\
&             & 0 &1/2& 0 & 1/2 & 0   & 1/2 & 0   & 1/2 & 0    & $S$\\
&             & 1   & 1, 2& 0 & 1, 4& 3   & 2, 5& 0   & 2, 7& 5 & $k$\\
\cline{2-12}
&             &4.0&3.0    &2.0&2.629&1.411&2.171&1.117&1.838&$0.938^{\dagger\dagger}$& $\Delta E/|J|$\\
&$\frac{3}{2}$& 3 & 16    & 3 & 8   & 3   & 8   & 3   &  8  & 3 & deg\\
&             & 1 &3/2    & 1 & 3/2 & 1   & 3/2 & 1   & 3/2 & 1 & $S$\\
&             & 0 &0, 1, 2& 2 & 2, 3& 0   & 1, 6& 4   &3, 6 & 0 & $k$\\
\cline{1-12}
&   & 12 & 6 & 10 &8.456&9.722&9.045&9.630&$9.263^{\dagger\dagger}$&$9.590^{\dagger\dagger}$& $E_0/(NJ)$\\
&$2$& 1  & 1 & 1  & 1   & 1   & 1   & 1   & 1 & 1 &  deg\\
&   & 0 & 0 & 0 & 0     & 0   & 0   & 0   & 0 & 0 & $S$\\
&   & 0  & 0 & 0  & 0   & 0   & 0   & 0   & 0 & 0 & $k$\\
\cline{2-12}
&             &4.0&2.0    &2.0&1.922&1.394&1.652&1.091&$1.431^{\dagger\dagger}$&$0.906^{\dagger\dagger}$& $\Delta E/|J|$\\
&$2$          & 3 & 9     & 3 &  6  & 3   & 6   & 3   & 6   & 3 & deg\\
&             & 1 & 1     & 1 &  1  & 1   & 1   & 1   & 1   & 1 & $S$\\
&             & 1 &0, 1, 2& 2 & 2, 3& 3   &3, 4 & 4   & 4, 5& 5 & $k$\\
\cline{1-12}
&             & 17.5 & 8.5 & 15 &12.434&14.645&13.451&$14.528^{\dagger}$&$13.848^{\dagger\dagger}$&$14.475^{\dagger\dagger}$& $E_0/(NJ)$\\
&$\frac{5}{2}$& 1    & 4   & 1  & 4    & 1    & 4    & 1 & 4 & 1 &  deg\\
&             & 0    &1/2  & 0  & 1/2  & 0    & 1/2  & 0 &1/2& 0 & $S$\\
&             & 1    & 1, 2& 0  & 1,4  & 3    & 2, 5 & 0 &2, 7& 5 & $k$\\
\hline
\end{tabular}
\vspace*{5mm}
\end{center}
\caption[]{Properties of ground and first excited state of AF
Heisenberg rings for various $N$ and $s$ (continuation): ground state energy
$E_0$, gap $\Delta E$, degeneracy $deg$, total spin $S$ and
shift quantum number $k$.
$\dagger$ -- O.~Waldmann, private communication.
$\dagger\dagger$ -- projection method \cite{Man:RMP91}. 
}\label{magmol:T-2-1-B}
\end{table}

Nowadays exact diagonalization\index{exact diagonalization}
methods allow to evaluate eigenvalues and eigenvectors of
$\op{H}$ for small even and odd spin rings of various numbers
$N$ of spin sites and spin quantum numbers $s$ where the
interaction is given by antiferromagnetic nearest neighbor
exchange
\cite{BoF:PR64,BoJ:PRB83,FLM:PRB91,Kar94,BSS:JMMM00:B,Schnack:PRB00}.
Although Marshall-Peierls sign rule and the theorems of Lieb,
Schultz, and Mattis do not apply to non-bipartite rings,
i.~e. frustrated rings with odd $N$, it turns out that such rings
nevertheless show astonishing regularities\footnote{Work done
with Klaus B\"arwinkel and Heinz-J\"urgen Schmidt, Universit\"at
Osnabr\"uck, Germany.}.  Unifying the picture for even and odd
$N$, we find for the ground state without exception
\cite{BSS:JMMM00:B,Schnack:PRB00}:
\begin{enumerate}
\item The ground state belongs to the subspace ${\mathcal H}(S)$
with the smallest possible total spin quantum number $S$;
this is either $S=0$ for $N\!\cdot\!s$ integer, then the total magnetic quantum
number $M$ is also zero, or $S=1/2$ for $N\!\cdot\!s$ half integer, then
$M=\pm 1/2$.
\item If $N\!\cdot\!s$ is integer, then the ground state is non-degenerate.
\item If $N\!\cdot\!s$ is half integer, then the ground state is fourfold degenerate.
\item If $s$ is integer or $N\!\cdot\!s$ even, then the shift
quantum number is $k=0$.
\item If $s$ is half integer and $N\!\cdot\!s$ odd, then the shift
quantum number turns out to be $k=N/2$.
\item If $N\!\cdot\!s$ is half integer, then $k=\lfloor(N+1)/4\rfloor$
and $k=N-\lfloor(N+1)/4\rfloor$ is found.
$\lfloor(N+1)/4\rfloor$ symbolizes the
greatest integer less or equal to $(N+1)/4$.
\end{enumerate}
In the case of $s=1/2$ one knows the $k$-quantum numbers for all
$N$ via the Bethe ansatz \cite{FLM:PRB91,Kar94}, and for spin $s=1$
and even $N$ the $k$ quantum numbers are consistent with
Ref.~\cite{BoJ:PRB83}.  

It appears that for the properties of the first excited state
such rules do not hold in general, but only for ``high enough"
$N>5$ \cite{Schnack:PRB00}. Then, as can be anticipated from
tables \xref{magmol:T-2-1-A} and \xref{magmol:T-2-1-B}, we can conjecture that
\begin{itemize}
\item if $N$ is even, then the first excited state has $S=1$ and
is threefold degenerate, and
\item if $N$ is odd and the single particle spin is
half-integer, then the first excited state has $S=3/2$ and is
eightfold degenerate, whereas
\item if $N$ is odd and the single particle spin is
integer, then the first excited state has $S=1$ and is
sixfold degenerate.
\end{itemize}

Considering relative ground states in subspaces ${\mathcal
H}(M)$ one also finds -- for even as well as for odd $N$ -- that
the shift quantum numbers $k$ show a strikingly simple
regularity for $N\ne 3$
\begin{eqnarray}
\label{magmol-E-4-37}
k
\equiv
\pm
(Ns-M)
\lceil
\frac{N}{2}
\rceil
\mod N
\ ,
\end{eqnarray}
where $\lceil N/2\rceil$ denotes the smallest integer greater
than or equal to $N/2$ \cite{BHS:PRB03}.  For $N=3$ and $3s-2\ge |M| \ge
1$ one finds besides the ordinary $k$-quantum numbers given by
\fmref{magmol-E-4-37} extraordinary $k$-quantum numbers, which
supplement the ordinary ones to the complete set
$\{k\}=\{0,1,2\}$.

For even $N$ the $k$ values form an alternating sequence
$0,N/2,0,N/2,\dots$ on descending from the magnon vacuum with
$M=Ns$ as known from the sign-rule of Marshall and Peierls
\cite{Mar:PRS55}. For odd $N$ it happens that the ordinary
$k$-numbers are repeated on descending from $M\le Ns-1$ to $M-1$
iff $N$ divides $[2(Ns-M)+1]$.

Using the $k$-rule one can as well derive a rule for the
relative ground state energies and for the respective $S$
quantum numbers:
\begin{itemize}
\item For the relative ground state energies one finds that
  if the $k$-number is different in adjacent subspaces,
  $E_{\text{min}}(S)<E_{\text{min}}(S+1)$ holds. If the
  $k$-number is the same, the energies could as well be the
  same.
\item Therefore, if $N$ (even or odd) does not divide
  $(2(Ns-M)+1) \lceil N/2\rceil$, then any relative ground state
  in ${\mathcal H}(M)$ has the total spin quantum number
  $S=|M|$.
\item This is always true for the absolute ground state which
  therefore has $S=0$ for $Ns$ integer and $S=1/2$ for $Ns$ half
  integer.
\end{itemize}

The $k$-rule \fmref{magmol-E-4-37} is founded in a
mathematically rigorous way for $N$ even
\cite{Mar:PRS55,LSM:AP61,LiM:JMP62}, $N=3$, $M=Ns$, $M=Ns-1$,
and $M=Ns-2$ \cite{BHS:PRB03}. An asymptotic proof for large enough
$N$ can be provided for systems with an asymptotically finite
excitation gap, i.~e. systems with integer spin $s$ for which
the Haldane conjecture applies \cite{Hal:PL83,Hal:PRL83}. In all
other cases numerical evidence was collected and the $k$-rule as
a conjecture still remains a challenge \cite{BHS:PRB03}.

\subsubsection*{Rotational bands}
\label{magmol-sec:4-3-2}

For many spin systems with constant isotropic antiferromagnetic
nearest neighbor Heisenberg exchange\index{Heisenberg model} the
minimal energies $E_{min}(S)$ form a rotational band,
i.~e. depend approximately quadratically on the total spin
quantum number $S$ \cite{ScL:PRB01,SLM:EPL01,Wal:PRB02}
\begin{eqnarray}
\label{magmol-E-4-42}
E_{min}(S)
\approx
E_a
- J\, \frac{D(N,s)}{N}\,
S (S+1)
\ .
\end{eqnarray}
The occurrence of a rotational band has been noted on several
occasions for an even number of spins defining a ring structure,
e.~g. see Ref.~\cite{Wal:PRB02}.  The minimal energies have been
described as ``following the Land\'{e} interval rule"
\cite{TDP:JACS94,LGC:PRB97A,LGC:PRB97B,ACC:ICA00}.  However,
we\footnote{Work done together with Marshall Luban, Ames Lab,
Iowa, USA.}  find that the same property also occurs for rings
with an odd number of spins as well as for the various polytope
configurations we have investigated, in particular for quantum
spins positioned on the vertices of a tetrahedron, cube,
octahedron, icosahedron, triangular prism, and an axially
truncated icosahedron.  Rotational modes have also been found in
the context of finite square and triangular lattices of spin-$1/2$
Heisenberg antiferromagnets \cite{BLL:PRB94,GSS:PRB89}.

\begin{figure}[ht!]
\centering
\includegraphics[height=35mm]{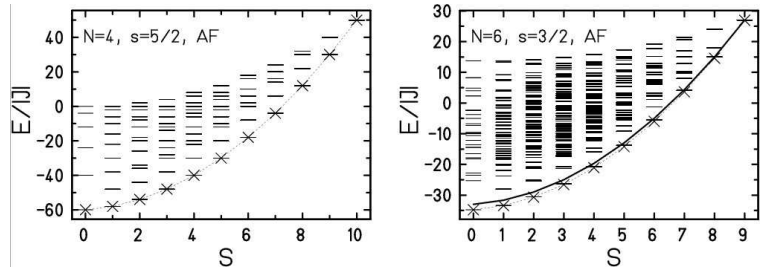}
\caption{Energy spectra of antiferromagnetically coupled
Heisenberg spin rings (horizontal dashes).  The crosses
connected by the dashed line represent the fit to the rotational
band according to \eqref{magmol-E-4-41}, which matches both the
lowest and the highest energies exactly.  On the l.h.s the
dashed line reproduces the exact rotational band, whereas on the
r.h.s. it only approximates it, but to high accuracy. The solid
line on the r.h.s. corresponds to the approximation
Eq.~\fmref{magmol-E-4-43}.}
\label{magmol-fig-rb-1}
\end{figure}

There are several systems, like spin dimers, trimers, squares,
tetrahedra, and octahedra which possess a strict rotational band
since their Hamiltonian can be simplified by quadrature.  As an
example the Heisenberg square, i.~e., a ring with $N=4$ is
presented. Because the Hamilton operator \fmref{magmol-E-4-18}
can be rewritten as
\begin{eqnarray}
\label{magmol-E-4-40}
\op{H}
&=&
-
J\,
\left(
\op{\vec{S}}^2 - \op{\vec{S}}_{13}^2 -\op{\vec{S}}_{24}^2
\right)
\ , \
\\
\op{\vec{S}}_{13}&=&\op{\vec{s}}(1)+\op{\vec{s}}(3)
\ , \
\op{\vec{S}}_{24}=\op{\vec{s}}(2)+\op{\vec{s}}(4)
\ ,
\end{eqnarray}
with all spin operators $\op{\vec{S}}^2$, $\op{\vec{S}}_{13}^2$
and $\op{\vec{S}}_{24}^2$ commuting with each other and with
$\op{H}$, one can directly obtain the complete set of
eigenenergies, and these are characterized by the quantum numbers
$S$, $S_{13}$ and $S_{24}$. In particular, the lowest energy for
a given total spin quantum number $S$ occurs for the choice
$S_{13}=S_{24}=2s$
\begin{eqnarray}
\label{magmol-E-4-41}
E_{min}(S)
=
-
J\,
\left[
S\,(S+1) - 2\cdot 2 s\, (2s+1)
\right]
=
E_0 - J\,S\,(S+1)
\ ,
\end{eqnarray}
where $E_0=4 s (2s+1) J$ is the exact ground state energy.  The
various energies $E_{min}(S)$ form a rigorous parabolic
rotational band of excitation energies.  Therefore, these
energies coincide with a parabolic fit (crosses connected by the
dashed line on the l.h.s. of Fig.~\ref{magmol-fig-rb-1}) passing
through the antiferromagnetic ground state energy and the
highest energy level, i.~e., the ground state energy of the
corresponding ferromagnetically coupled system.

It turns out that an accurate formula for the coefficient
$D(N,s)$ of \eqref{magmol-E-4-41} can be developed using the
sublattice structure of the spin array \cite{ScL:PRB01}. As an
example we repeat the basic ideas for Heisenberg rings with an
even number of spin sites \cite{ACC:ICA00}.  Such rings are
bipartite and can be decomposed into two sublattices, labeled
$A$ and $B$, with every second spin belonging to the same
sublattice.  The classical ground state (N\'eel state) is given
by an alternating sequence of opposite spin directions. On each
sublattice the spins are mutually parallel.  Therefore, a
quantum trial state, where the individual spins on each
sublattice are coupled to their maximum values, $S_A=S_B=N s /
2$, could be expected to provide a reasonable approximation to
the true ground state, especially if $s$ assumes large
values. For rings with even $N$ the approximation to the
respective minimal energies for each value of the total spin
$\op{\vec{S}}=\op{\vec{S}}_A + \op{\vec{S}}_B$ is then given by
\cite{ACC:ICA00}
\begin{eqnarray}
\label{magmol-E-4-43}
E_{min}^{\mbox{\scriptsize approx}}(S)
=
- \frac{4\, J}{N}\,
\left[
S (S+1) - 2 \frac{N s}{2} \left( \frac{N s}{2} + 1 \right)
\right]
\ .
\end{eqnarray}
This approximation exactly reproduces the energy of the highest
energy eigenvalue, i.~e., the ground state energy of the
corresponding ferromagnetically coupled system ($S=N s$). For
all smaller $S$ the approximate minimal energy
$E_{min}^{\mbox{\scriptsize approx}}(S)$ is bounded from below
by the true one (Rayleigh-Ritz variational principle).  The
solid curve displays this behavior for the example of $N=6$,
$s=3/2$ in Fig.~\ref{magmol-fig-rb-1} (r.h.s.).  The coefficient
``4" in Eq.~\fmref{magmol-E-4-43} is the classical value,
i.~e. for each fixed even $N$ the coefficient $D(N,s)$ approaches
4 with increasing $s$ \cite{ScL:PRB01}.

The approximate spectrum, \eqref{magmol-E-4-43}, is similar to
that of two spins, $\op{\vec{S}}_A$ and $\op{\vec{S}}_B$, each
of spin quantum number $N s / 2$, that are coupled by an
effective interaction of strength $4 J/N$. Therefore, one can
equally well say, that the approximate rotational band
considered in \fmref{magmol-E-4-43} is associated with an
effective Hamilton operator
\begin{eqnarray}
\label{magmol-E-4-44}
\op{H}^{\mbox{\scriptsize approx}}
&=&
- \frac{4\, J}{N}\,
\left[
\op{\vec{S}}^2 - \op{\vec{S}}_A^2 - \op{\vec{S}}_B^2
\right]
\ ,
\end{eqnarray}
where the two sublattice spins, $\op{\vec{S}}_A,
\op{\vec{S}}_B$, assume their maximal value $S_A=S_B=N s /
2$. Hamiltonian \fmref{magmol-E-4-44} is also known as
Hamiltonian of the Lieb-Mattis model which describes a system
where each spin of one sublattice interacts with every spin of
the other sublattice with equal strength
\cite{LiM:JMP62,Ric:PRB93}. 

\begin{figure}[ht!]
\centering
\includegraphics[height=50mm]{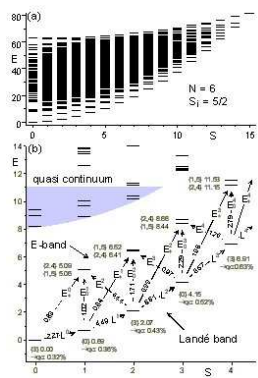}
\caption{The low-lying levels of a spin ring, $N=6$ and $s=5/2$
  in this example, can be grouped into the lowest
  (\textbf{L}and\'e) band, the first excited
  (\textbf{E}xcitation) band and the quasi-continuum (QC). For
  the spin levels of the L- and E-band $k$ is given in brackets
  followed by the energy. Arrows indicate strong transitions
  from the L-band. Associated numbers give the total oscillator
  strength $f_0$ for these transitions. With friendly
  permission by Oliver Waldmann \cite{Wal:PRB02}.} 
\label{magmol-fig-waldmann-bands}
\end{figure}
It is worth noting that this Hamiltonian reproduces more than
the lowest levels in each subspace ${\mathcal H}(S)$. At least
for bipartite systems also a second band is accurately
reproduced as well as the gap to the quasi-continuum above,
compare Figure~\ref{magmol-fig-waldmann-bands} and
Ref.~\cite{Wal:PRB02}.  This property is very useful since the
approximate Hamiltonian allows the computation of several
observables without diagonalizing the full Hamiltonian.

It is of course of utmost importance whether the band structure
given by the approximate Hamiltonian \fmref{magmol-E-4-44}
persists in the case of frustrated molecules. It seems that at
least the minimal energies still form a rotational band which is
understandable at least for larger spin quantum numbers $s$
taking into account that the parabolic dependence of the minimal
energies on $S$ mainly reflects the classical limit for a wide
class of spin systems \cite{ScL:JPA03}.

The following example demonstrates that even in the case of the
highly frustrated molecule \mofe\ the minimal energies arrange
as a ``rotational band''\footnote{Work done with Matthias Exler,
Universit\"at Osnabr\"uck, Germany.}.  In the case of \mofe\ the
spin system is decomposable into three sub-lattices with
sub-lattice spin quantum numbers $S_A$, $S_B$, and $S_C$
\cite{ScL:PRB01,SLM:EPL01}. The corresponding approximate
Hamilton operator reads
\begin{equation}
  \label{magmol-E-4-45}
  \op{H}_{\rm approx}=-J\frac{D}{N}\left[\op{\vec{S}}^2
    -\gamma\left(\op{\vec{S}}^2_A
      +\op{\vec{S}}^2_B+\op{\vec{S}}^2_C\right)\right]
\ ,
\end{equation}
where $\op{\vec{S}}$ is the total spin operator and the others
are sub-lattice spin operators. $D$ and $\gamma$ are allowed to
deviate from their respective classical values, $D=6$ and
$\gamma=1$, in order to correct for finite $s$.

\begin{figure}[ht!]
\centering
\includegraphics[clip,height=50mm]{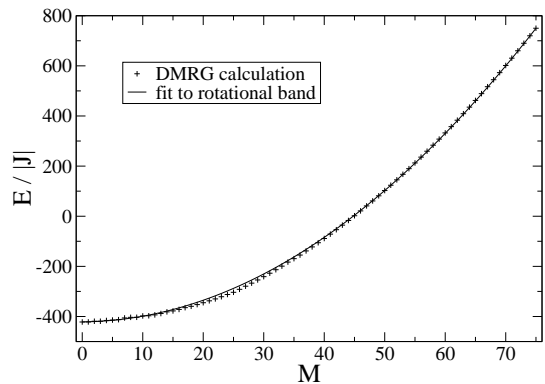}
\caption{DMRG eigenvalues and lowest
rotational band of the $s=5/2$ icosidodecahedron; $m=60$ was
used except for the lowest and first exited level which were
calculated with $m=120$.} 
\label{magmol-fig-6}
\end{figure}
\index{DMRG}
We use the DMRG method to approximate the lowest energy
eigenvalues of the full Hamiltonian and compare them to those
predicted by the rotational band hypothesis
\fmref{magmol-E-4-45}.  Fig.~\ref{magmol-fig-6} shows the
results and a fit to the lowest rotational band. Assuming the
same dependence on $m$ as in the $s=1/2$ case, the relative
error of the DMRG data should also be less than $1\%$.  The
agreement between the DMRG energy levels and the predicted
quadratic dependence is very good. Nevertheless, it remains an
open question whether higher lying bands are present in such a
highly frustrated compound.

\subsubsection*{Magnetization jumps}
\label{magmol-sec:4-3-4}

\begin{figure}[ht!]
\centering
\includegraphics[height=35mm]{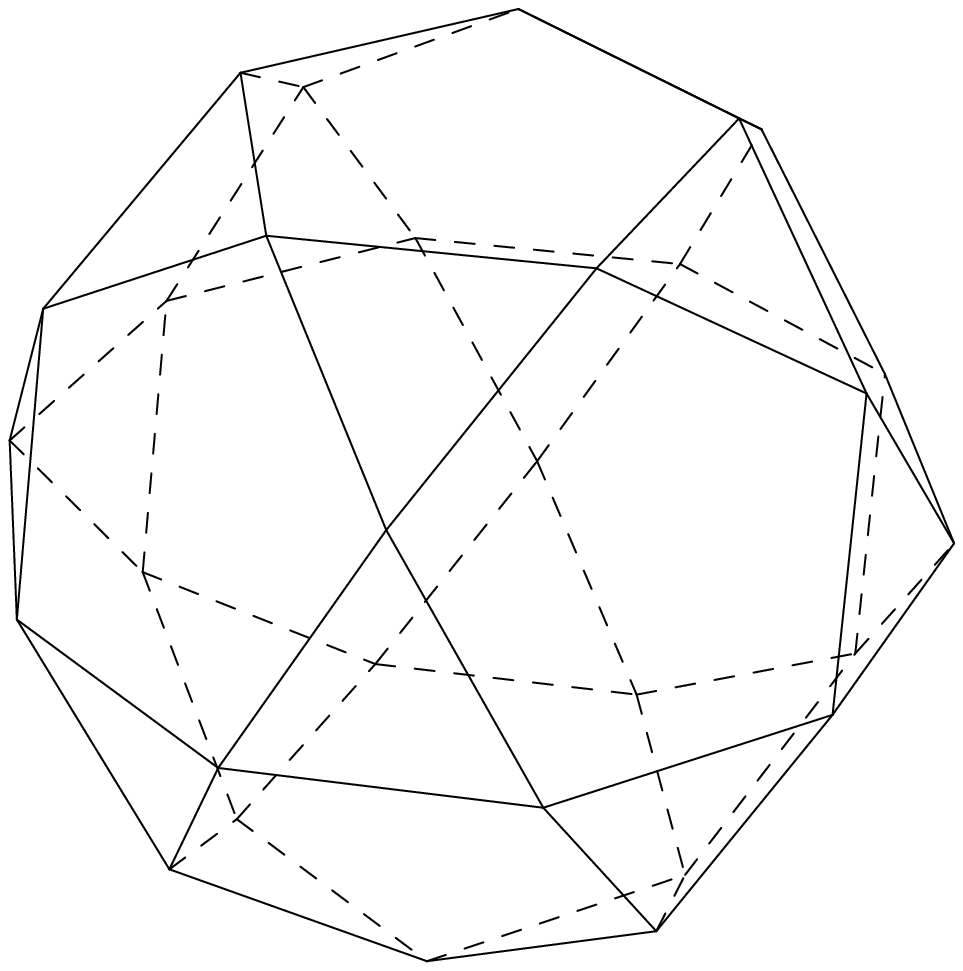}
$\qquad$
\includegraphics[height=35mm]{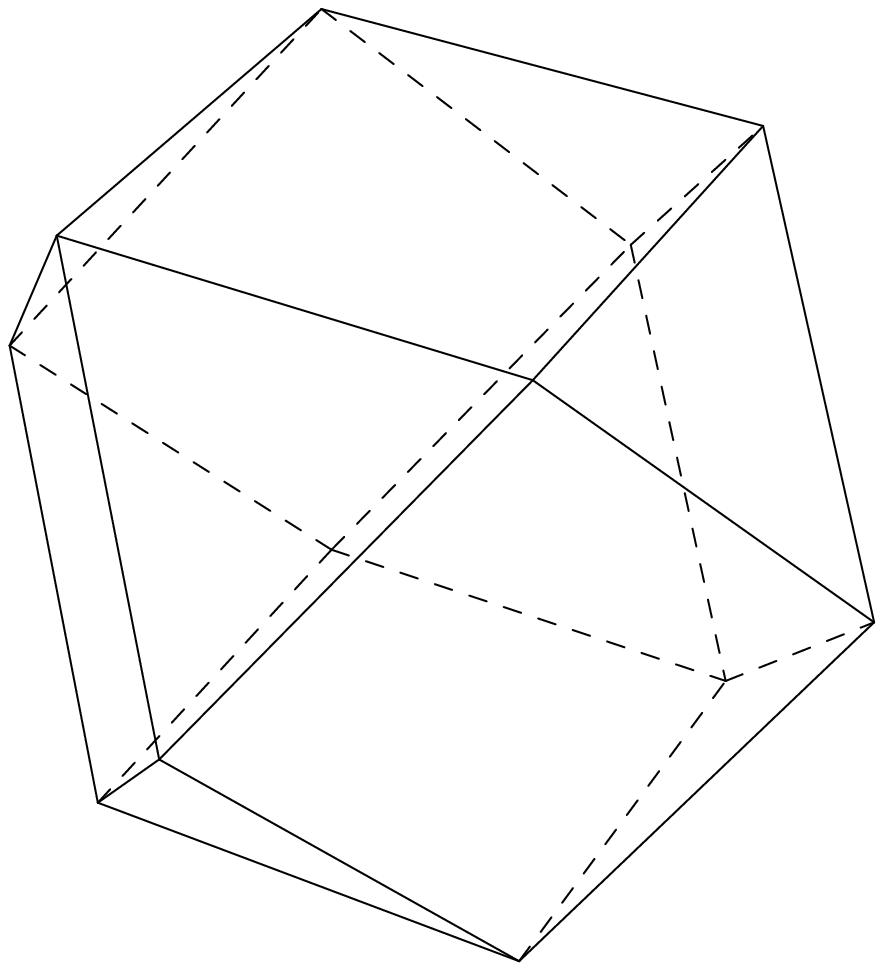}
\caption{Structure of the icosidodecahedron (l.h.s.) and the
  cuboctahedron (r.h.s.).}
\label{magmol-fig-octahedra}
\end{figure}
Although the spectra of many magnetic molecules possess a
rotational band of minimal energies $E_{min}(S)$ and although in
the classical limit, where the single-spin quantum number $s$
goes to infinity, the function $E_{min}(S)$ is even an exact
parabola if the system has co-planar ground states
\cite{ScL:JPA03}, we\footnote{Work done with Heinz-J\"urgen
Schmidt, Universit\"at Osnabr\"uck, Andreas Honecker,
Universit\"at Braunschweig, Johannes Richter and J\"org
Schulenburg, Universit\"at Magdeburg, Germany.} find that for
certain coupling topologies, including the cuboctahedron and the
icosidodecahedron (see Fig.~\ref{magmol-fig-octahedra}), that
this rule is violated for high total spins
\cite{SSR:EPJB01,SHS:PRL02}.  More precisely, for the
icosidodecahedron the last four points of the graph of $E_{min}$
versus $S$, i.~e.~the points with $S=S_{max}$ to $S=S_{max}-3$,
lie on a straight line
\begin{eqnarray}
\label{magmol-E-4-46}
E_{min}(S)
&=&
60 J s^2
-
6 J s (30 s - S)
\ .
\end{eqnarray}
An analogous statement holds for the last three points of the
corresponding graph for the cuboctahedron. These findings are
based on numerical calculations of the minimal energies for
several $s$ both for the icosidodecahedron as well as for the
cuboctahedron.  For both and other systems a rigorous proof of
the high spin anomaly can be given
\cite{SSR:EPJB01,Schmidt:JPA02}. 

The idea of the proof can be summarized as follows: A necessary
condition for the anomaly is certainly that the minimal energy
in the one-magnon space is degenerate.  Therefore, localized
one-magnon states can be constructed which are also of minimal
energy. When placing a second localized magnon on the spin array
there will be a chance that it does not interact with the first
one if a large enough separation can be achieved. This new
two-magnon state is likely the state of minimal energy in the
two-magnon Hilbert space because for antiferromagnetic
interaction two-magnon bound states do not exist. This procedure
can be continued until no further independent magnon can be
placed on the spin array. In a sense the system behaves as if it
consists of non-interacting bosons which, up to a limiting
number, can condense into a single-particle ground state.  In
more mathematical terms: In order to prove the high-spin anomaly
one first shows an inequality which says that all points
$(S,E_{min}(S))$ lie above or on the line connecting the last
two points.  For specific systems as those mentioned above what
remains to be done is to construct particular states which
exactly assume the values of $E_{min}$ corresponding to the
points lying on the bounding line, then these states are
automatically states of minimal energy.

\begin{figure}[ht!]
\centering
\includegraphics[height=35mm]{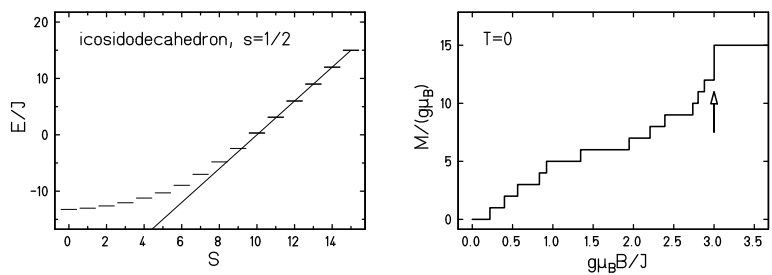}
\caption{Icosidodecahedron: L.h.s. -- minimal energy levels
$E_{min}(S)$ as a function of total spin $S$. R.h.s. --
magnetization curve at $T=0$ \cite{SSR:EPJB01}.} 
\label{magmol-fig-independent-magnons}
\end{figure}
The observed anomaly -- linear instead of parabolic dependence
-- results in a corresponding jump of the magnetization curve
${\mathcal M}$ versus $B$, see
Fig.~\ref{magmol-fig-independent-magnons}. In contrast, for
systems which obey the Land\'{e} interval rule the magnetization
curve at very low temperatures is a staircase with equal steps
up to the highest magnetization.  The anomaly could indeed be
observed in magnetization measurements of the Keplerate
molecules \{Mo$_{72}$Fe$_{30}$\}.  Unfortunately, the
magnetization measurements \cite{MLS:CPC01,SLM:EPL01} performed
so far suffer from too high temperatures which smear out the
anomaly.

Nevertheless, it may be possible to observe truly giant
magnetization jumps in certain two-dimensional spin systems
which possess a suitable coupling (e.~g. Kagom\'e)
\cite{SHS:PRL02}. In such systems the magnetization jump can be
of the same order as the number of spins, i.~e. the jump remains
finite -- or in other words is macroscopic -- in the
thermodynamic limit $N\rightarrow\infty$. Thus, this effect is a
true macroscopic quantum effect.

\section{Dynamics}
\label{magmol-sec:5}

In this section I would like to outline two branches --
tunneling and relaxation -- where the dynamics of magnetic
molecules is investigated. The section is kept rather
introductory since the field is rapidly evolving and it is too
early to draw a final picture on all the details of the involved
processes.

\subsection{Tunneling}
\label{magmol-sec:5-1}

\index{magnetization tunneling}\index{tunneling}
Tunneling dynamics has been one of the corner stones in
molecular magnetism since its very early days, see
e.~g.~\cite{FST:PRL96,TLB:Nature96,ThB:JLTP98,THB:JMMM99,FPC:Nature00}.

The subject can roughly be divided into two parts, one deals
with tunneling processes of the magnetization in molecules
possessing a high ground state spin and an anisotropy barrier,
the second deals with the remaining tunneling processes,
e.~g. in molecules which have an $S=0$ ground state.

\begin{figure}[ht!]
\centering
\includegraphics[height=38mm]{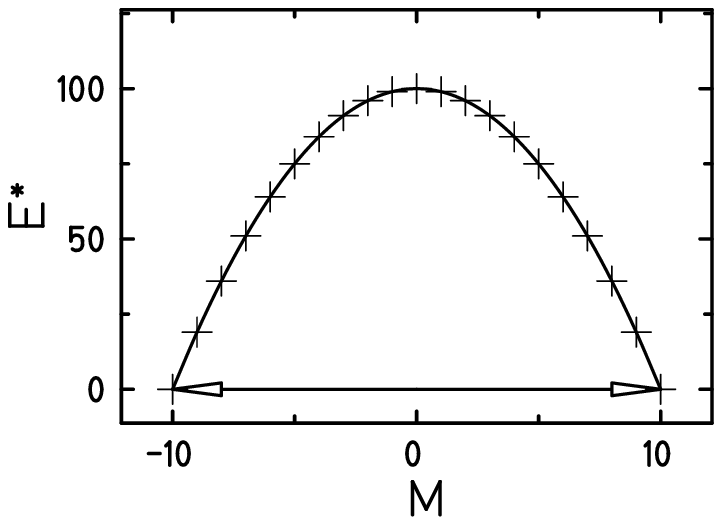}
$\qquad$
\includegraphics[height=38mm]{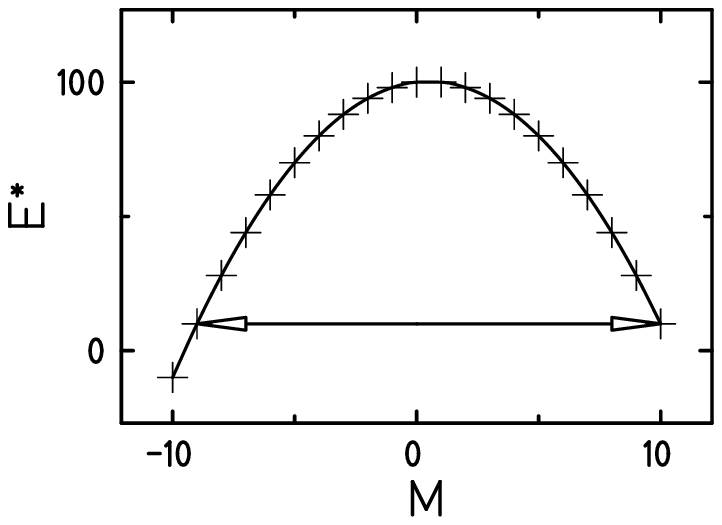}
\caption{Sketch of the tunneling barrier for a high spin
  molecule with $S=10$, l.h.s. without magnetic field,
  r.h.s. with magnetic field, compare
  Eq.~\fmref{magmol-E-4-1}. The arrows indicate a possible 
  resonant tunneling process.}
\label{magmol-fig-tunnel}
\end{figure}
As already mentioned in section \ref{magmol-sec:4-1-1} some
molecules like Mn$_{12}$ and Fe$_{8}$ possess a high ground
state spin. Since the higher lying levels are well separated
from the low-lying $S=10$ levels a single-spin Hamiltonian
\fmref{magmol-E-4-1}, which includes an anisotropy term, is
appropriate. Figure~\ref{magmol-fig-tunnel} sketches the energy
landscape for an anisotropy term which is quadratic in
$\op{S}_z$. If the Hamiltonian includes terms like a magnetic
field in $x$-direction that do not commute with $\op{S}_z$
resonant tunneling is observed between states $\ket{S,M}$ and
$\ket{S,-M}$. This behavior is depicted on the l.h.s. of
Fig.~\ref{magmol-fig-tunnel} for the transition between $M=-10$
and $M=10$. If an additional magnetic field is applied in
$z$-direction the quadratic barrier acquires an additional
linear Zeeman term and is changed like depicted on the r.h.s. of
Fig.~\ref{magmol-fig-tunnel}. Now tunneling is possible between
states of different $|M|$, see e.~g.~\cite{BCG:JPSJ00}.

It is rather simple to model the tunneling process in the model
Hilbert space of $S=10$. i.~e. a space with dimension
$2S+1=21$. Nevertheless, in a real substance the tunneling
process is accompanied and modified by other influences. The
first major factor is temperature which may enhance the process,
this leads to thermally assisted tunneling \cite{LTB:JAP97}.
Each such substance hosts phonons which modify the tunneling
process, too, resulting in phonon assisted tunneling
\cite{PRH:PRL95,BPS:PRL96,PRH:PRL96,CGJ:PRL00}. Then local
dipolar fields and nuclear hyperfine fields may strongly affect
the relaxation in the tunneling regime \cite{CGS:JMMM99}. In
addition there may be topological quenching due to the symmetry
of the material \cite{Garg:PRB01A,Garg:PRB01B,PG:PRB02}.  And
last but not least describing such complicated molecules not in
effective single-spin models but in many-spin models is still in
an unsatisfactory state, compare \cite{RHD:JMMM02}.

\begin{figure}[ht!]
\centering
\includegraphics[height=20mm]{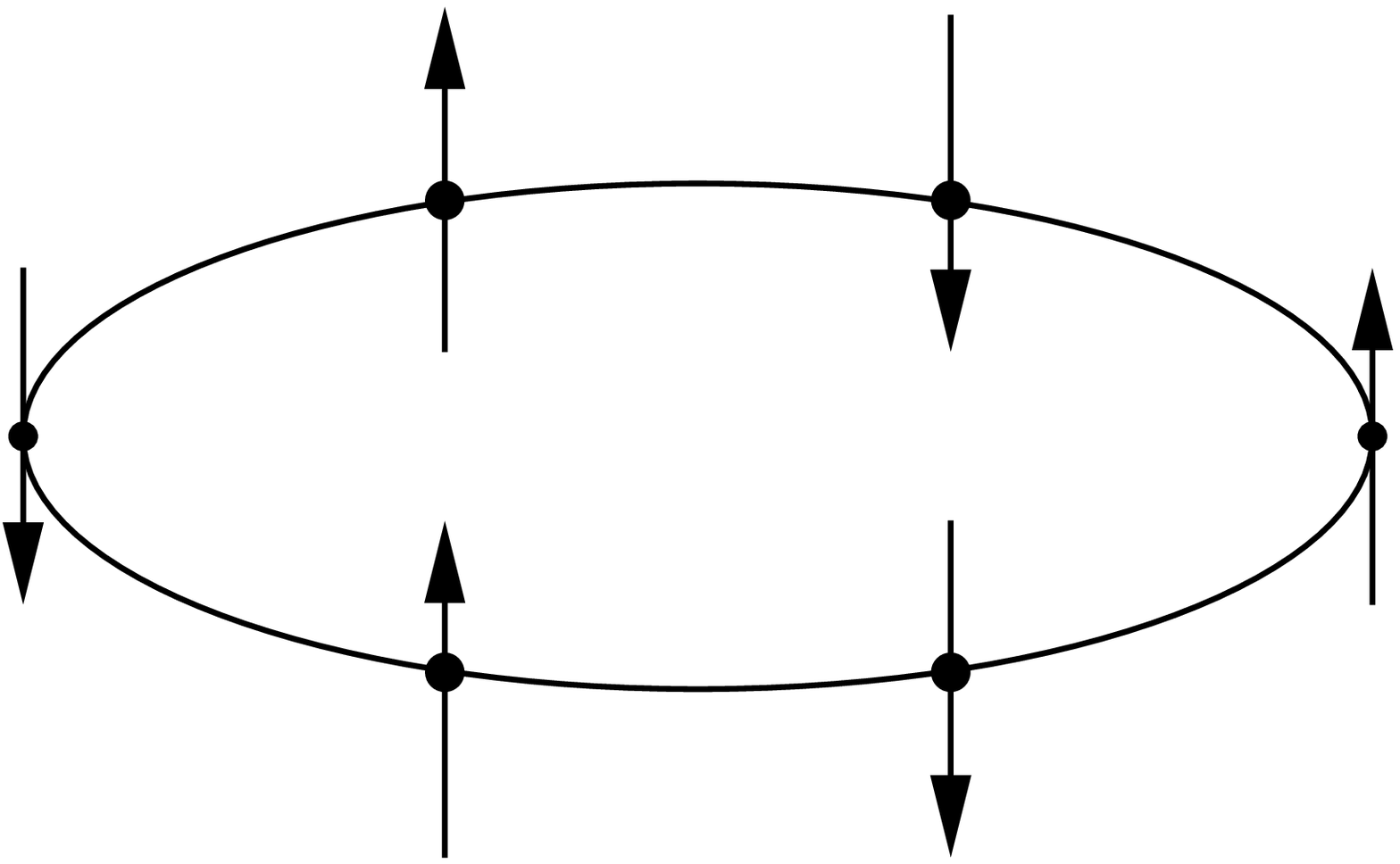}
\raisebox{8mm}{$\qquad\Leftrightarrow\qquad$}
\includegraphics[height=20mm]{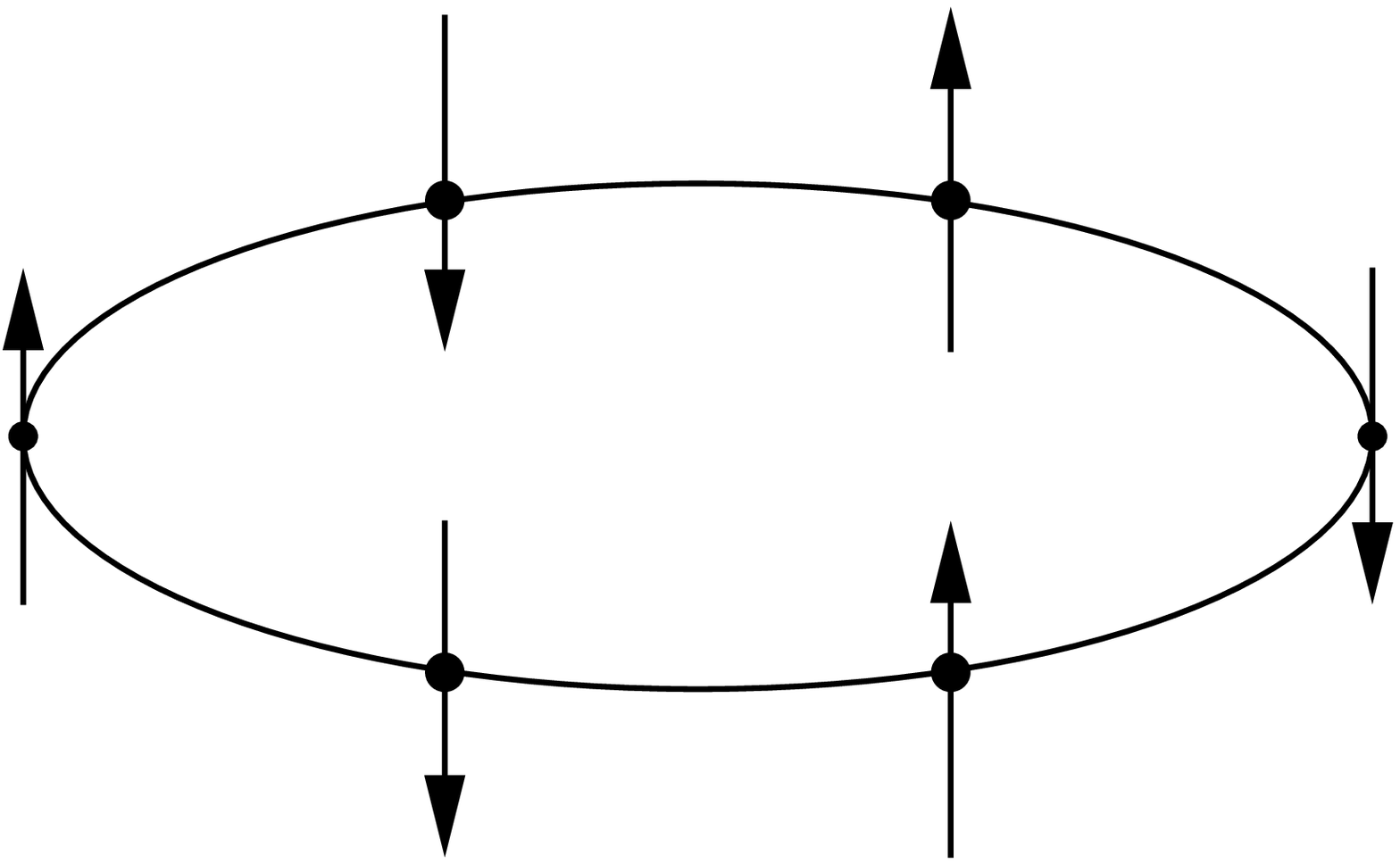}
\caption{Sketch of the tunneling process between \Neel-like
  states on a spin ring. Without loss of generality the state on
  the l.h.s. will be denoted by $\ket{\text{\Neel}, 1}$ and the state on
  the r.h.s. will be denoted by $\ket{\text{\Neel}, 2}$.}
\label{magmol-fig-sc}
\end{figure}
Another kind of tunneling is considered for Heisenberg spin
rings with uniaxial single-ion anisotropy. Classically the
ground state of even rings like Na:Fe$_{6}$ and Cs:Fe$_{8}$ is
given by a sequence of spin up and down like in
Fig.~\ref{magmol-fig-sc}. It now turns out that such a \Neel-like
state, which is formulated in terms of spin-coherent states
\fmref{magmol-E-4-35}, contributes dominantly to the true ground
state as well as to the first excited state if the anisotropy is
large enough \cite{HML:EPJB02}. Thus it is found that the ground
state $\ket{E_0}$ and the first excited state $\ket{E_1}$ can be
approximated as
\begin{eqnarray}
\label{magmol-E-4-61}
\ket{E_0}
&\approx&
\frac{1}{\sqrt{2}}
\left(
\ket{\text{\Neel}, 1} \pm \ket{\text{\Neel}, 2}
\right)
\\
\ket{E_1}
&\approx&
\frac{1}{\sqrt{2}}
\left(
\ket{\text{\Neel}, 1} \mp \ket{\text{\Neel}, 2}
\right)
\nonumber
\ ,
\end{eqnarray}
where the upper sign is appropriate for rings where the number
of spins $N$ is a multiple of 4, e.~g. $N=8$, and the lower sign
is for all other even $N$.

Therefore, the tunneling frequency is approximately given by the
gap between ground and first excited state. Experimentally, such
a tunnel process is hard to observe, especially since ESR is
sensitive only to the total spin. What would be needed is a
local probe like NMR. This could be accomplished by replacing
one of the iron ions by another isotope.

\begin{figure}[ht!]
\centering
\includegraphics[height=45mm]{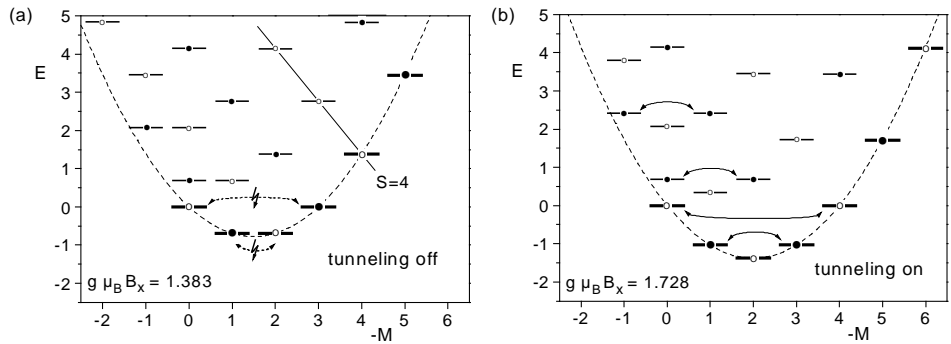}
\caption{Energy spectrum of spin rings with $N=6$ and vanishing
  anisotropy at two magnetic fields drawn as a function of the
  magnetic quantum number $M$. The dashed curves represent the
  lowest-lying parabolas $E_{min}(M)$ discussed in
  section~\ref{magmol-sec:4-3-2}. A white or black circle
  indicates that a state belongs to $k=0$ or $k=N/2$, compare
  Fig.~\ref{magmol-fig-waldmann-bands}. States belonging to one
  spin multiplet are located on straight lines like that plotted
  in panel (a) for the $S=4$ multiplet. With friendly
  permissions by Oliver Waldmann \cite{Waldmann:EPL02}.}
\label{magmol-fig-tunnel-waldmann}
\end{figure}
The tunneling process was further analyzed for various values of
the uniaxial single-ion anisotropy \cite{Waldmann:EPL02}. Since
in such a case the cyclic shift symmetry persists, $k$ is still
a good quantum number. Therefore, mixing of states is only
allowed between states with the same $k$ quantum number. This
leads to the conclusion that the low-temperature tunneling
phenomena can be understood as the tunneling of the spin vector
between different rotational modes with $\Delta S = 2$, compare
Fig.~\ref{magmol-fig-tunnel-waldmann} and the subsection on
rotational bands on page~\pageref{magmol-sec:4-3-2}.

\subsection{Relaxation dynamics}
\label{magmol-sec:5-2}

In a time-dependent magnetic field the magnetization tries to
follow the field. Looking at this process from a microscopic
point of view, one realizes that, if the Hamiltonian would
commute with the Zeeman term, no transitions would occur, and
the magnetization would not change a tiny bit. There are
basically two sources which permit transitions: non-commuting
parts in the spin Hamiltonian and interactions with the
surrounding. In the latter case the interaction with phonons
seems to be most important.

Since a complete diagonalization of the full Hamiltonian
including non-commuting terms as well as interactions like the
spin-phonon interaction is practically impossible, both
phenomena are modeled with the help of rate equations.

\begin{figure}[ht!]
\centering
\includegraphics[height=35mm]{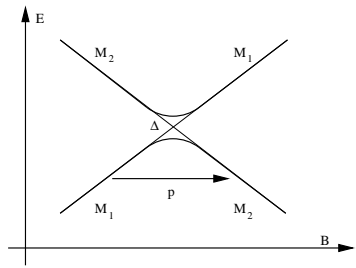}
\caption{Schematic energy spectrum in the vicinity of an avoided
  level crossing. The formula by Landau, Zener, and
  St\"uckelberg \fmref{magmol-E-4-62} approximates the
  probability $p$ for the tunneling process from $M_1$ to
  $M_2$. }
\label{magmol-fig-lzs}
\end{figure}
If we start with a Hamiltonian $\op{H}_0$, which may be the
Heisenberg Hamiltonian and add a non-commuting term
$\op{H}^\prime$ like anisotropy the eigenstates of the full
Hamiltonian are superpositions of those of $\op{H}_0$. This
expresses itself in avoided level crossings where the spectrum
of $\op{H}_0$ would show level crossings, compare
Fig.~\ref{magmol-fig-lzs}. Transitions between eigenstates of
$\op{H}_0$, which may have good $M$ quantum numbers, can then
effectively be modeled with the formula by Landau, Zener, and
St\"uckelberg
\cite{Lan:PZS32,Zen:PRS32,Stueckelberg:HPA32,RMS:PRB97,WSC:JAP00}
\begin{eqnarray}
\label{magmol-E-4-62}
p
&=&
1 - \exp\left\{
-\frac{\pi \Delta^2}{2 \hbar g \mu_B |M_1-M_2|\; \ddt B}
\right\}
\ .
\end{eqnarray}
$\Delta$ denotes the energy gap at the avoided level crossing. 

The effect of phonons is taken into account by means of two
principles: detailed balance, which models the desire of the
system to reach thermal equilibrium and energy conservation,
which takes into account that the energy released or absorbed by
the spin system must be absorbed or released by the phonon
system and finally exchanged with the thermostat. The
interesting effects arise since the number of phonons is very
limited at temperatures in the Kelvin-range or below, thus they
may easily be used up after a short time (phonon bottleneck) and
have to be provided by the thermostat around which needs a
characteristic relaxation time. Since this all happens in a
time-dependent magnetic field, the Zeeman splittings change all
the time and phonons of different frequency are involved at each
time step. In addition the temperature of the spin system
changes during the process because the equilibration with the
thermostat (liq. Helium) is not instantaneous. More accurately
the process is not in equilibrium at all, especially for
multi-level spin systems. Only for two-level systems the
time-dependent occupation can be translated into an apparent
temperature. In essence the retarded dynamics leads to distinct
hysteresis loops which have the shape of a butterfly
\cite{WAH:Nature02,SLC:CPL02,WKS:PRL02}. For more detailed
information on dissipative two-level systems the interested
reader is referred to Ref.~\cite{LCD:RMP87,WKS:PRL02}.

\section{Magnetocalorics}
\label{magmol-sec:6}

The mean (internal) energy, the magnetization and the magnetic
field are thermodynamic observables just as pressure and volume.
Therefore, we can design thermodynamic processes which work with
magnetic materials as a medium. This has of course already been
done for a long time. The most prominent application is
magnetization cooling which is mainly used to reach sub-Kelvin
temperatures \cite{GiM:PR33}. The first observation of sub-Kelvin
temperatures is a nice example of how short an article can be to
win the Nobel prize (Giauque, Chemistry, 1949). Meanwhile
magnetization cooling is used in ordinary refrigerators.

\begin{figure}[ht!]
\centering
\includegraphics[height=60mm]{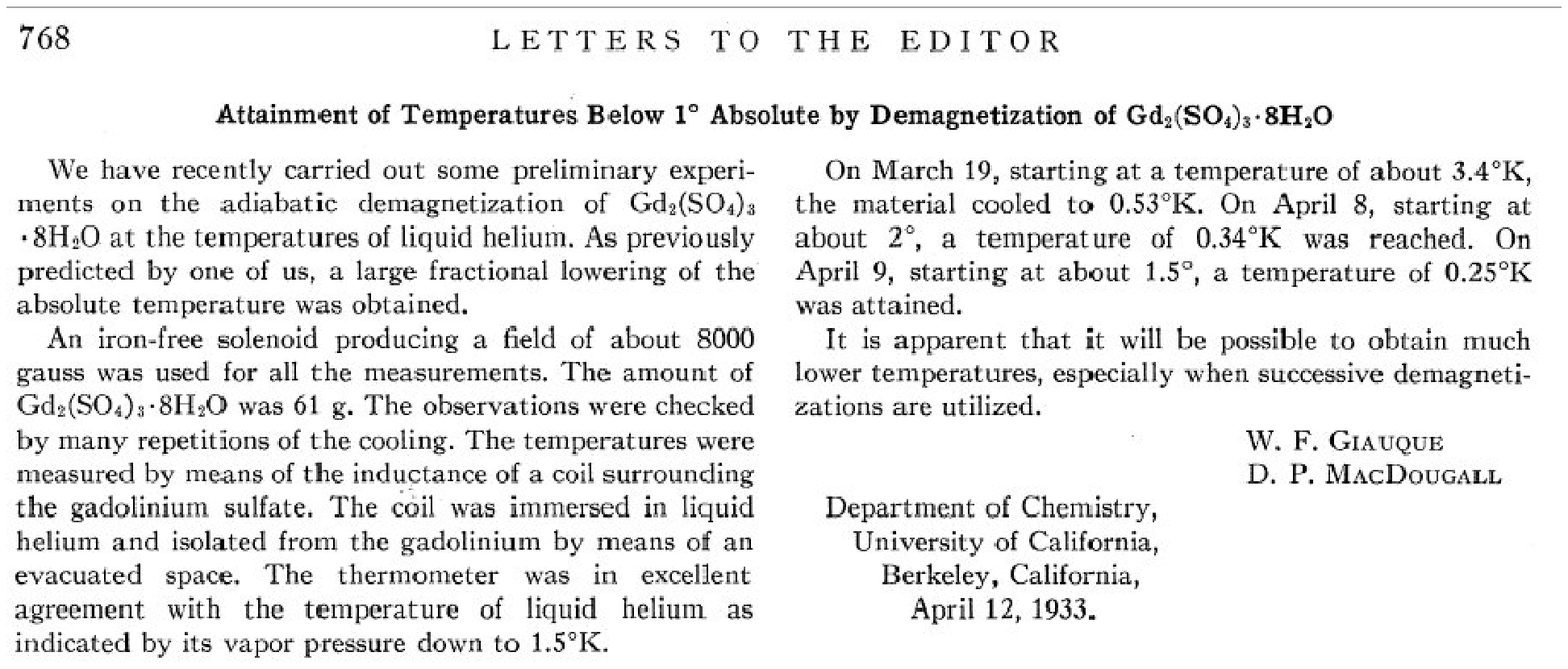}
\caption{The first observation of sub-Kelvin temperatures
  \cite{GiM:PR33} is a nice example of how short an article can
  be to win the Nobel prize (Giauque, Chemistry, 1949).}
\label{magmol-fig-subK}
\end{figure}

In early magnetocaloric experiments simple refrigerants like
paramagnetic salts have been used. We will therefore consider
such examples first.

For a paramagnet the Hamiltonian consists of the Zeeman term
only. We then obtain for the partition function
\begin{eqnarray}
\label{magmol-E-6-01}
Z(T,B,N)
&=&
\left\{
\frac{\text{sinh}[\beta g \mu_B B (s+1/2)]}
     {\text{sinh}[\beta g \mu_B B/2]}
\right\}^N
\ .
\end{eqnarray}
Then the magnetization is
\begin{eqnarray}
\label{magmol-E-6-02}
{\mathcal M}(T,B,N)
&=&
N g \mu_B
\left\{
(s+1/2)\text{coth}[\beta g \mu_B B (s+1/2)]
-1/2 \text{sinh}[\beta g \mu_B B/2]
\right\}
\ ,
\end{eqnarray}
and the entropy reads
\begin{eqnarray}
\label{magmol-E-6-03}
S(T,B,N)
&=&
N k_B
\text{ln}\left\{
\frac{\text{sinh}[\beta g \mu_B B (s+1/2)]}
     {\text{sinh}[\beta g \mu_B B/2]}
\right\}
-
k_B \beta B {\mathcal M}(T,B,N)
\ .
\end{eqnarray}
Besides their statistical definition both quantities follow from
the general thermodynamic relationship
\begin{eqnarray}
\label{magmol-E-6-04}
\dint F
&=&
\left(
\pp{F}{T}
\right)_B
\dint T
+ 
\left(
\pp{F}{B}
\right)_T
\dint B
=
-S \dint T
-
{\mathcal M}
\dint B
\ ,
\end{eqnarray}
where $F(T,B,N)=-k_B T \; \text{ln}[Z(T,B,N)]$.

Looking at Eq.~\fmref{magmol-E-6-01} it is obvious that all
thermodynamic observables for a paramagnet depend on temperature
and field via the combination $B/T$, and so does the
entropy. Therefore, an adiabatic demagnetization ($S=const$)
means that the ratio $B/T$ has to remain constant, and thus
temperature shrinks linearly with field, i.e.
\begin{eqnarray}
\label{magmol-E-6-05}
\left(
\pp{T}{B}
\right)_S^{\text{para}}
&=&
\frac{T}{B}
\ .
\end{eqnarray}
This situation changes completely for an interacting spin system.
Depending on the interactions the adiabatic cooling rate
$\pp{T}{B}$ can be smaller or bigger than the paramagnetic one
\fmref{magmol-E-6-05} and even change sign, i.e. one would
observe heating during demagnetization. It is nowadays
understood that the cooling rate acquires extreme values close to
phase transitions due to the excess entropy associated with such
processes \cite{Zhi:PRB03,ZGR:PRL03,ZhH:JSM04,DeR:PRB04}.

\begin{figure}[ht!]
\centering
\includegraphics[width=35mm]{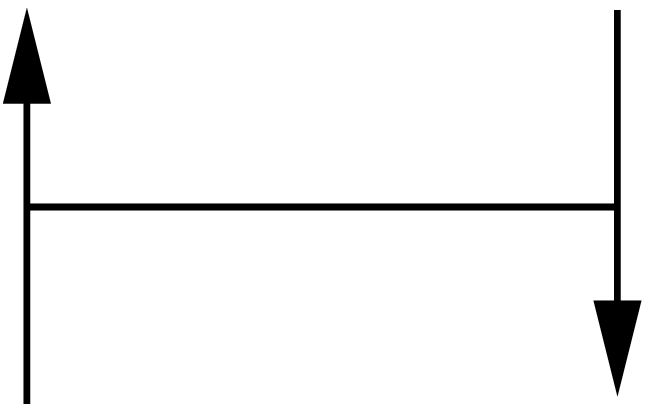}
\qquad\qquad
\includegraphics[width=45mm]{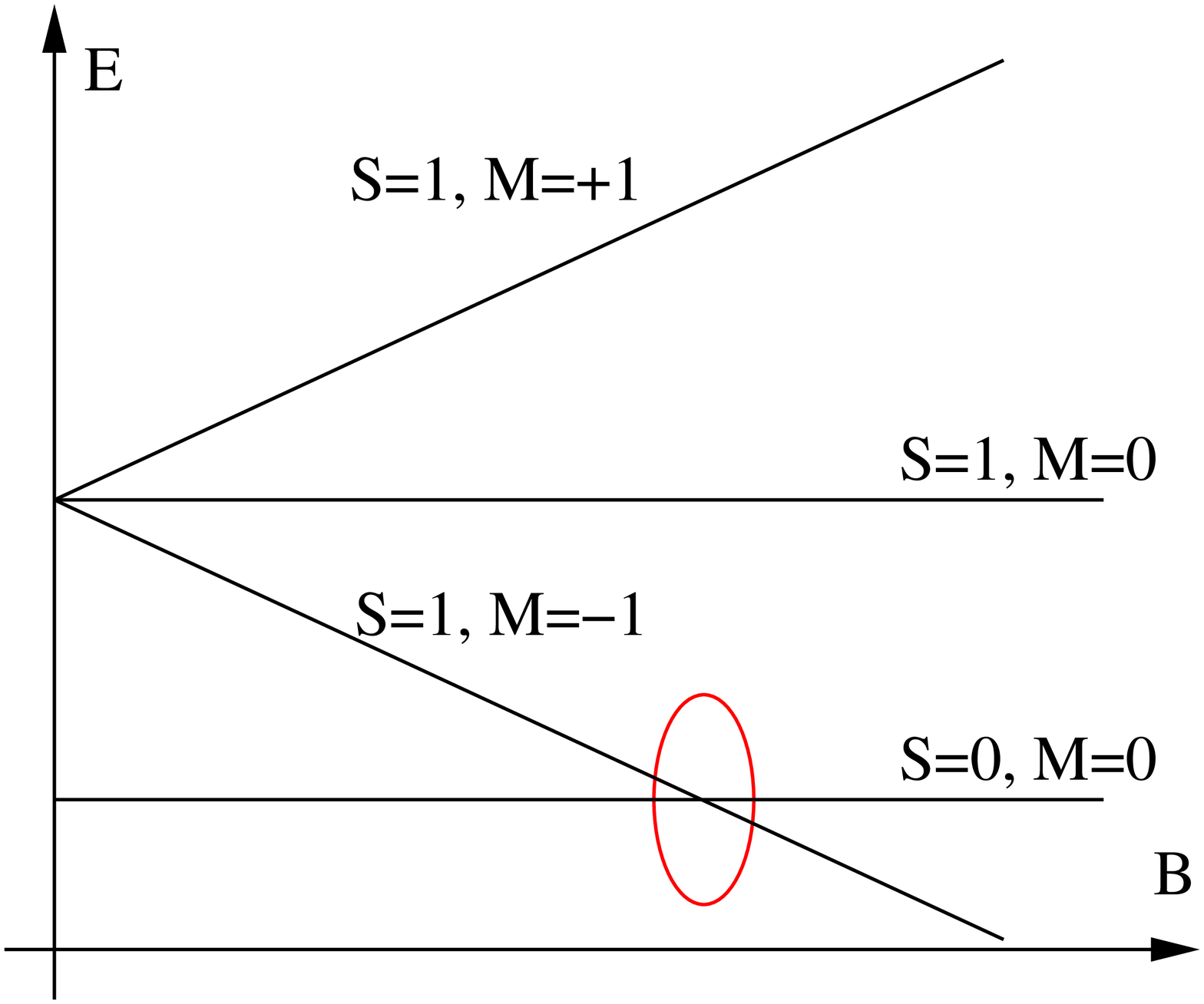}
\caption{L.h.s.: Sketch of an antiferromagnetically coupled spin
  dimer. R.h.s.: Dependence of the energy levels on magnetic
  field for an antiferromagnetically coupled spin-$1/2$
  dimer. At the critical field $B_c$ the lowest triplet level
  crosses the ground state level with $S=0$.}
\label{magmol-fig-d1}
\end{figure}

In the following this statement will be made clear by discussing
the example of a simple antiferromagnetically coupled spin-$1/2$
dimer (Fig.~\xref{magmol-fig-d1}, l.h.s.). In a magnetic field such
a system experiences a ``quantum phase transition'' if the
lowest triplet level crosses the original ground state with
$S=0$, see Fig.~\xref{magmol-fig-d1}, r.h.s.. Although one would
hesitate to call such an ordinary ground state level crossing
quantum phase transition it nevertheless is one. At $T=0$ the
magnetization ${\mathcal M}(T=0,B)$ is a non-analytic function
of the magnetic field $B$. At the critical field $B_c$ where the
levels cross the magnetization exhibits a step.

\begin{figure}[ht!]
\centering
\includegraphics[width=65mm]{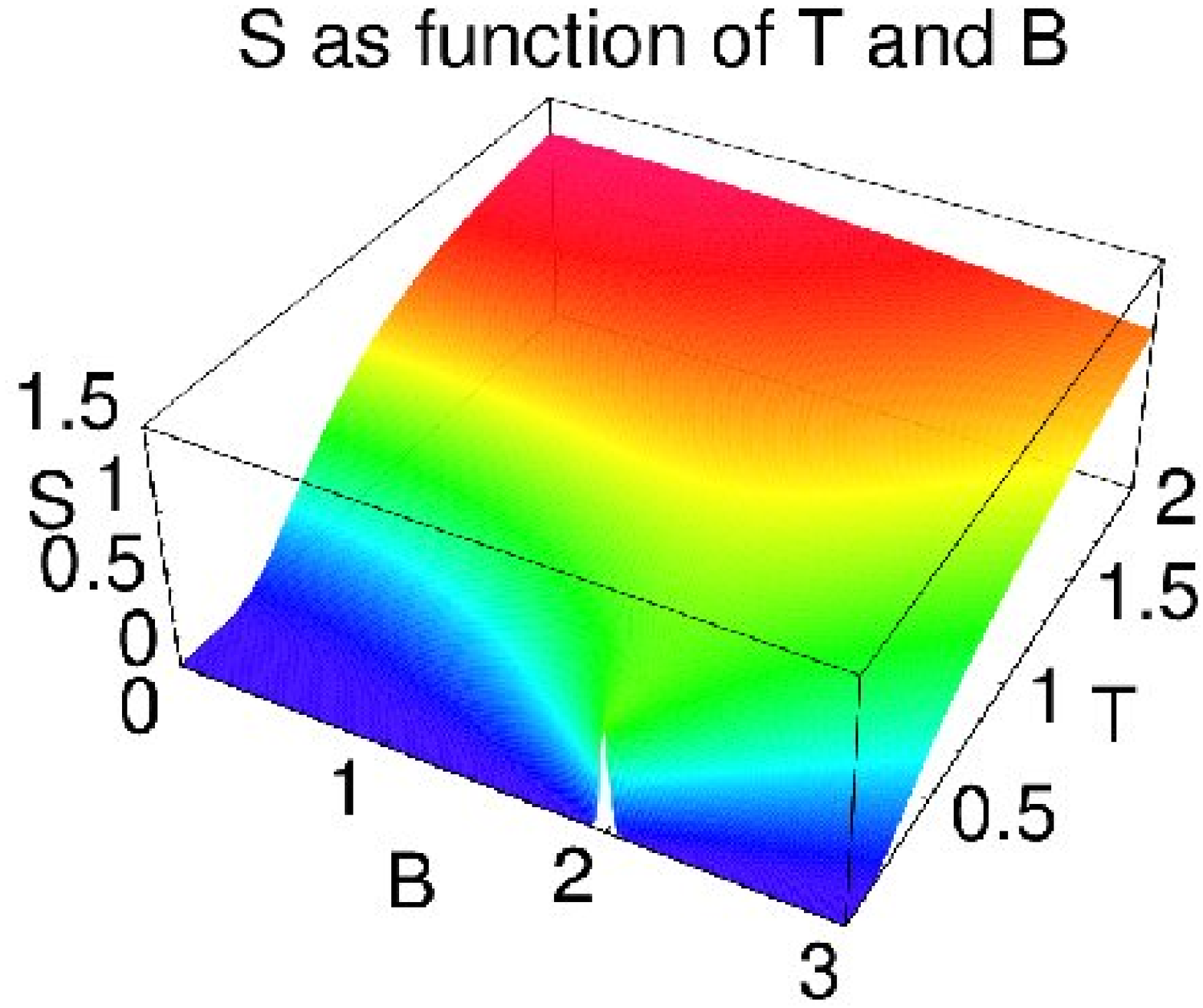}
\qquad
\includegraphics[width=65mm]{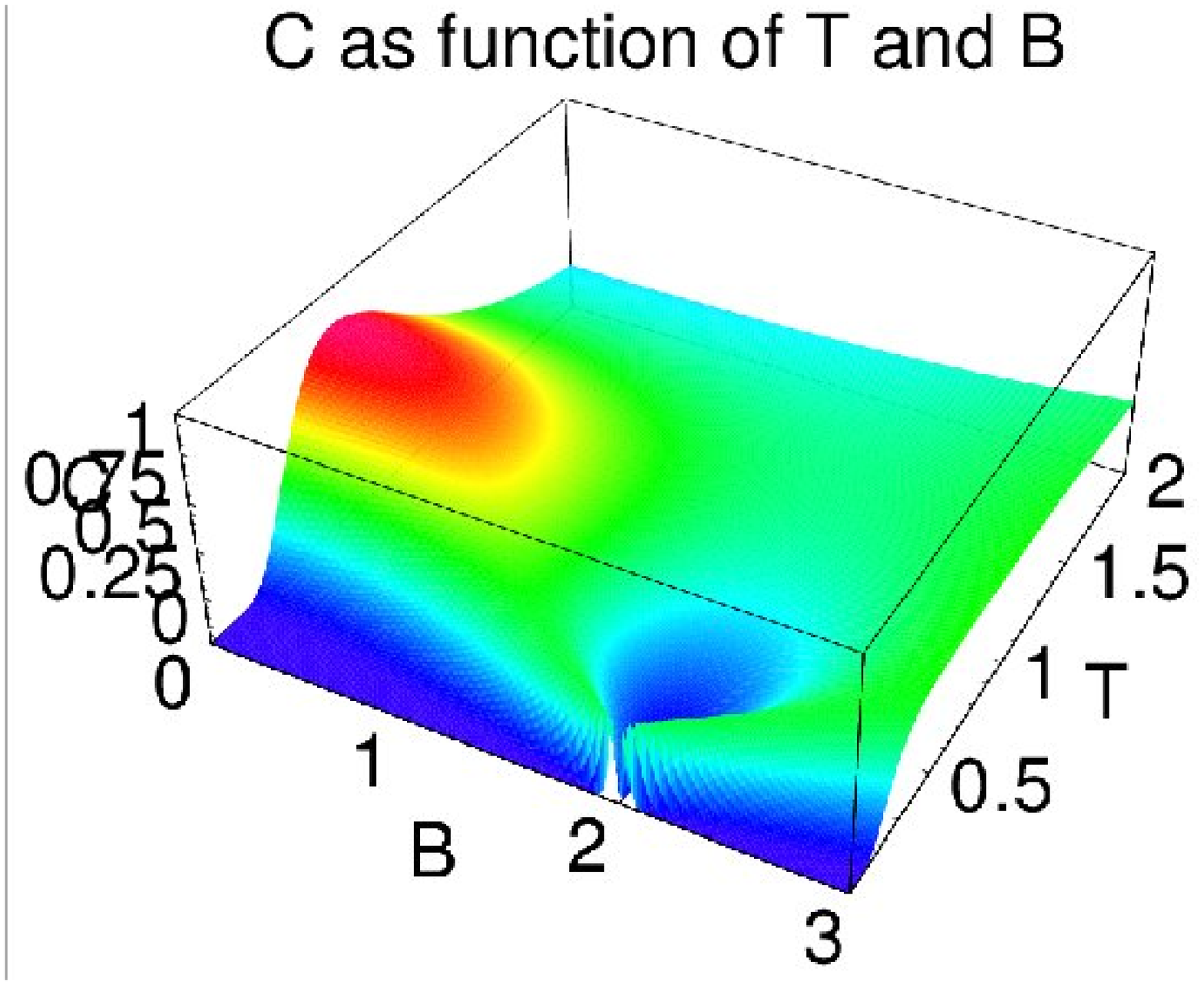}
\caption{L.h.s.: Entropy of the dimer
  (Fig.~\xref{magmol-fig-d1}) as function of $B$ and $T$.
  R.h.s.: Heat capacity of the dimer as function of $B$ and
  $T$.}
\label{magmol-fig-d2}
\end{figure}

In addition the entropy, which at $T=0$ is zero for the
non-degenerate ground state acquires a finite value at the
critical field $B_c$ due to the degeneracy of the crossing
levels. This enhancement remains present even at temperatures
$T>0$, compare Fig.~\xref{magmol-fig-d2}, l.h.s.. In addition
the heat capacity varies strongly around the critical field as
is shown in Fig.~\xref{magmol-fig-d2}, r.h.s..

\begin{figure}[ht!]
\centering
\includegraphics[width=65mm]{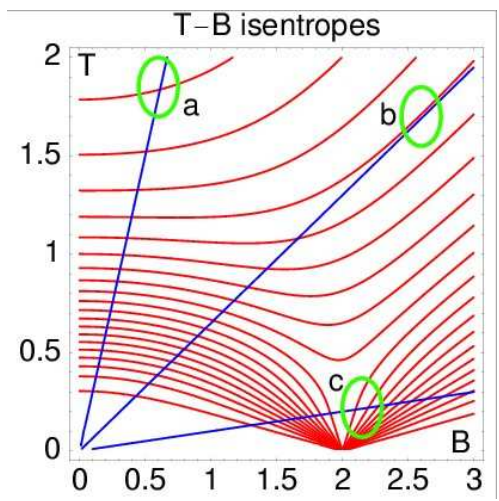}
\caption{Isentrops of the spin dimer. The straight lines show
  the behavior of a paramagnet for comparison. $B$ is along the
  $x$-axis, $T$ along the $y$-axis.}
\label{magmol-fig-d3}
\end{figure}

The behavior of the entropy as well as of the heat capacity
explains how the adiabatic cooling rate 
\begin{eqnarray}
\label{magmol-E-6-06}
\left(
\pp{T}{B}
\right)_S
=
-T
\frac{\left(\pp{S}{B}\right)_T}{C}
\end{eqnarray}
depends on field and temperature. Figure~\xref{magmol-fig-d3}
shows the isentropes of the antiferromagnetically coupled dimer
both as function of field $B$ and temperature $T$.  The straight
lines show the behavior of a paramagnet for comparison. Three
regions are highlighted. 
\begin{itemize}
\item a: For low fields and high temperatures $\pp{T}{B}$ is
  smaller than for the paramagnet.
\item b: For high fields and high temperatures the interacting
  system assumes the paramagnetic limit, i.e. $\pp{T}{B}$ is the
  same in both systems.
\item c: For low temperatures and fields just above the critical
  field $\pp{T}{B}$ is much bigger than the cooling rate
  of the paramagnet.
\item Not highlighted but nevertheless very exciting is the
  region at low temperature just below the critical field where
  the ``cooling'' rate $\pp{T}{B}$ has the opposite sign, i.e.
  upon demagnetizing the system heats up and upon magnetizing
  the system cools down.
\end{itemize}
The rate $\pp{T}{B}$ \fmref{magmol-E-6-06} depends directly on
the derivative of the entropy with respect to the magnetic
field. Therefore, it is clear that the effect will be enhanced
if a high degeneracy can be obtained at some critical field.
This is indeed possible in several frustrated materials where
giant magnetization jumps to saturation are observed
\cite{SHS:PRL02,RSH:JPCM03,ZhH:JSM04,DeR:PRB04}.

\begin{figure}[ht!]
\centering
\includegraphics[width=65mm]{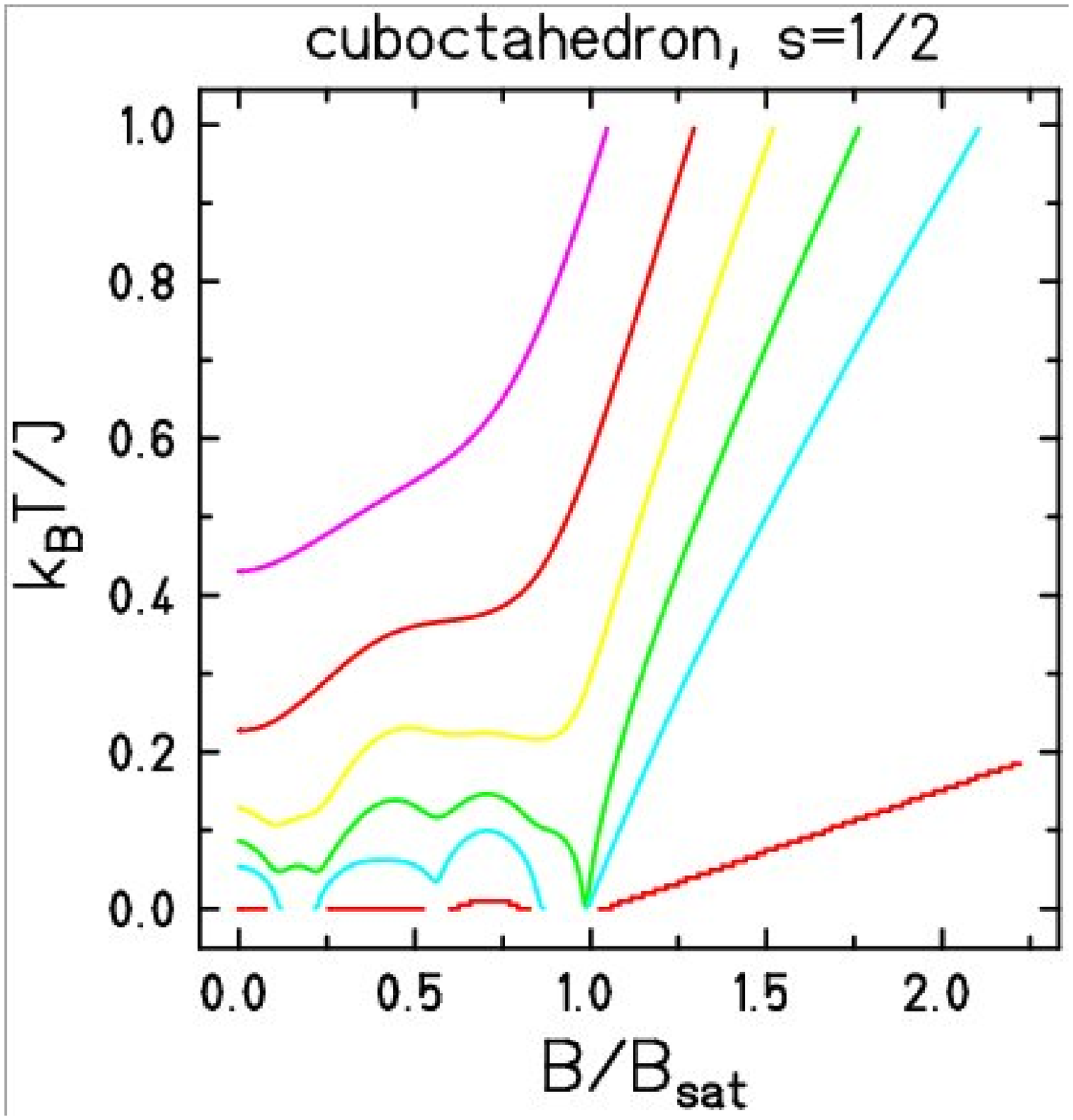}
\qquad
\includegraphics[width=65mm]{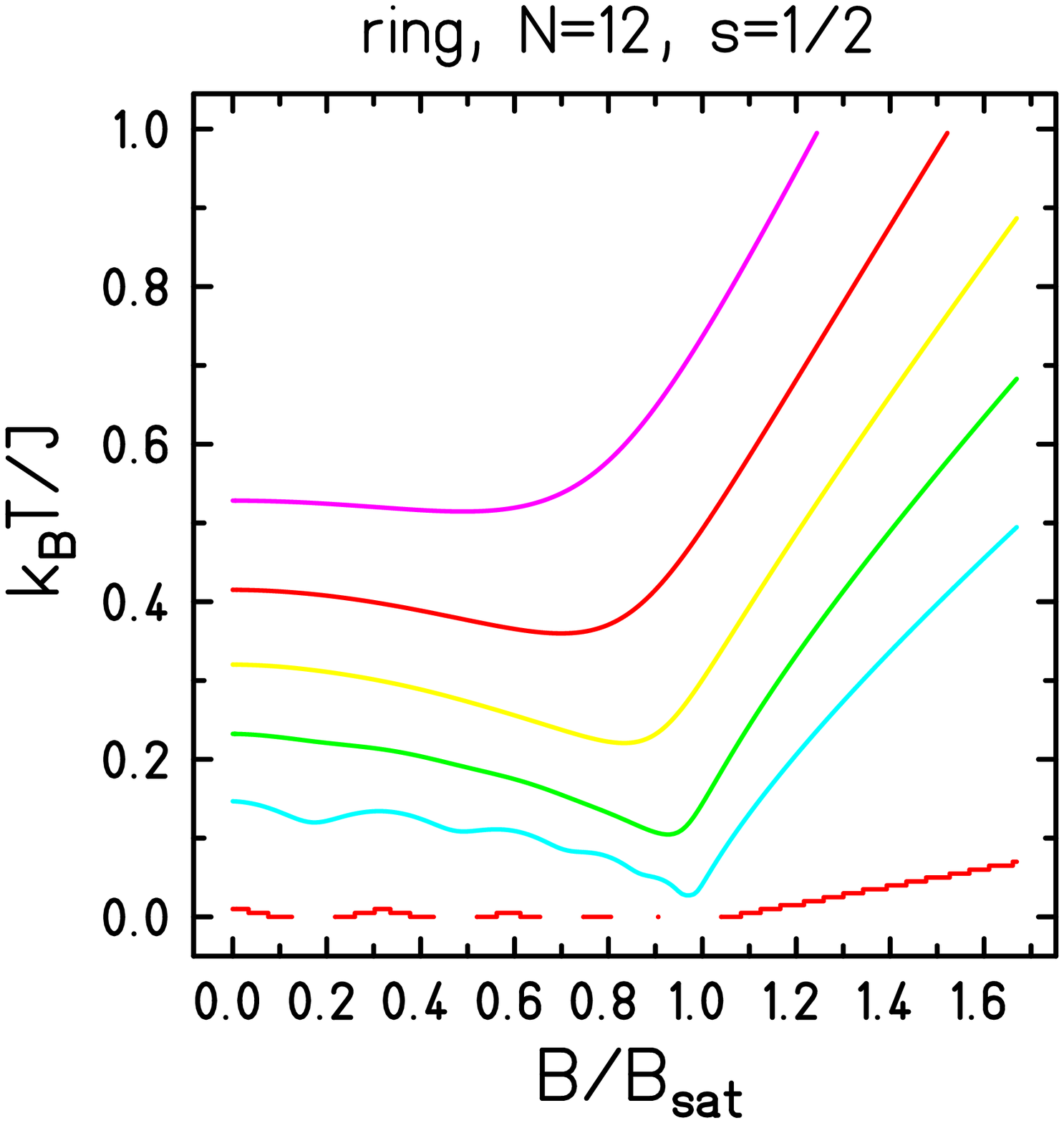}
\caption{Isentropes of an antiferromagnetically coupled
  spin-$1/2$ cuboctahedron (l.h.s.) in comparison to the
  isentropes of an antiferromagnetically coupled spin ring of 12
  spins with $s=1/2$ (r.h.s.).  The structure of the
  cuboctahedron is shown in Fig.~\xref{magmol-fig-octahedra} (r.h.s.).}
\label{magmol-fig-mce}
\end{figure}

The following Figure~\xref{magmol-fig-mce} compares isentropes
of an antiferromagnetically coupled spin-$1/2$ cuboctahedron
(l.h.s.) with the isentropes of an antiferromagnetically coupled
spin ring of 12 spins with $s=1/2$ (r.h.s.). It is clearly
visible that the cuboctahedral spin system has steeper
isentropes than the ring system, and the reason is that the
cuboctahedron features a larger jump to magnetization saturation
($\Delta M=2$) than the spin ring ($\Delta M=1$) which is
connected with an enhanced degeneracy and thus higher entropy.
 
Summarizing, one can say that low-dimensional frustrated spin
systems and especially magnetic molecules are substances with an
interesting magnetocaloric behavior and may turn out to be
useful new refrigerants for special applications.

\newpage
\begin{center}
\textbf{Acknowledgement}
\end{center}

I would like to thank my colleagues Klaus B\"arwinkel, Mirko
Br\"uger, Matthias Exler, Peter Hage, Frank Hesmer, Detlef
Mentrup, and Heinz-J\"urgen Schmidt at the university of
Osnabr\"uck as well as Paul K\"ogerler, Marshall Luban, Robert
Modler, and Christian Schr\"oder at the Ames lab, Ames, Iowa,
USA for the fruitful collaboration which produced many of the
discussed results.

I would also like to thank Jens Kortus, Oliver Waldmann, and
Wolfgang Wernsdorfer for pointing out valuable literature for
further reading.

I further like to thank Klaus B\"arwinkel, Peter Hage, and
Heinz-J\"urgen Schmidt for carefully reading the manuscript.

This work was supported by the Deutsche Forschungsgemeinschaft
(Grant No. SCHN~615/5-1 and SCHN~615/8-1).




\end{document}